\documentstyle[12pt]{article} 
\input{epsf.tex}

\begin{document}

\vskip -1cm

\begin{center} 
 
{\Large \bf Morris$\,$-Thorne wormholes with a \\ 
cosmological constant}\\ 
\vskip 0.2cm
{\bf Jos\'e P. S. Lemos}\\
{\footnotesize   Department of Physics, Columbia University,}\\
{\footnotesize New York, NY 10027,}$\,\,$
{\footnotesize \&}\\
{\footnotesize Centro Multidisciplinar de Astrof\'{\i}sica - CENTRA,} \\
{\footnotesize Departamento de F\'{\i}sica, Instituto Superior T\'ecnico,}\\
{\footnotesize  Av. Rovisco Pais 1, 1049-001 Lisbon}\\
{\footnotesize e-mail: lemos@physics.columbia.edu}\\
\vskip 0.1cm
{\bf Francisco S. N. Lobo} \\
{\footnotesize  Centro de Astronomia e Astrof\'{\i}sica 
da Universidade de Lisboa,} \\
{\footnotesize  Campo Grande, Ed. C8 1749-016 Lisbon}\\
{\footnotesize e-mail: flobo@cosmo.fis.fc.ul.pt}
\vskip 0.1cm
{\bf S\'ergio Quinet de Oliveira} \\
{\footnotesize  Observat\'orio Nacional - MCT,} \\
{\footnotesize Rua General Jos\'e Cristino 77,
20921-400 Rio de Janeiro}\\
{\footnotesize e-mail: quinet@hotmail.com}\\
\end{center}  
\begin{abstract}  
First, the ideas introduced in the wormhole research field since the
work of Morris and Thorne are reviewed, namely, the issues of
energy conditions, wormhole construction, stability, time machines and
astrophysical signatures.  Then, spherically symmetric and static
traversable Morris$\,$-Thorne wormholes in the presence of a generic
cosmological constant $\Lambda$ are analyzed. A matching of an
interior solution to the unique exterior vacuum solution is done using
directly the Einstein equations. The structure as well as several
physical properties and characteristics of traversable wormholes due
to the effects of the cosmological term are studied.  Interesting
equations appear in the process of matching.  For instance, one finds
that for asymptotically flat and anti-de Sitter spacetimes the surface
tangential pressure $\cal P$ of the thin shell, at the boundary of the
interior and exterior solutions, is always strictly positive, whereas
for de Sitter spacetime it can take either sign as one could expect,
being negative (tension) for relatively high $\Lambda$ and high
wormhole radius, positive for relatively high mass and small wormhole
radius, and zero in-between. Finally, some specific solutions with
$\Lambda$, based on the Morris$\,$-Thorne solutions, are provided.
 
\end{abstract} 
 
\newpage
 
\section{Review and Introduction}

It is now 15 years  that traversable 
wormhole theory started in earnest from the
work of Mike Morris and Kip Thorne 
published 
in 1988 \cite{Morris}. It was first introduced as a tool for teaching general
relativity, as well as an attempt to allure young students into the
field, for instance those that had read {\it Contact}, a novel of Carl
Sagan that uses a wormhole to shortcut 
a large astronomical distance, but it rapidly spread into several
branches.  These developments culminated with the publication of the
book {\it Lorentzian Wormholes: From Einstein to Hawking} by Visser
\cite{Visser}, where a review on the subject up to 1995, as well as new
ideas are developed and hinted at.  It is our intention in this introduction
to do a brief review on the subject of wormholes. The subject has grown
substantially, and it is now almost out of control. We will focus on
the work developed after Visser's book was published (most of the
references prior to its releasing are in it), 
paying attention to the issues that branched out of \cite{Morris},  
like the issue of energy conditions, wormhole construction, stability,
time machines and astrophysical signatures.

\subsection{The beginning} 

It is true that Wheeler \cite{wheeler1}, before the work of Morris and
Thorne \cite{Morris}, had seen wormholes, such as Reissner-Nordstr\"om
or Kerr wormholes, as objects of the quantum foam connecting different
regions of spacetime and operating at the Planck scale, which were
transformed later into Euclidean wormholes by Hawking \cite{hawking0}
and others, but these Wheeler wormholes were not traversable, one
could not cross them from one side to the other and back, and
furthermore would, in principle, develop some type of singularity
\cite{geroch}.  Having been a student of Wheeler, and having further
learned through Wheeler's interaction with Zel'dovich on the trace
energy condition (which states $\rho\geq3p$, with $\rho$ being the
energy density and $p$ the pressure of the fluid on its rest frame)
that energy conditions are on a shaky ground
\cite{wheelerredbook,zeldovichnovikovbook}, Thorne together with his
student Morris \cite{Morris}, understood that wormholes, with two
mouths and a throat, might be objects of nature, as stars and black
holes are.  Indeed, it is a basic fact for the construction of
traversable wormholes that the null energy condition, the weakest of
the conditions, has to be violated.

\subsection{Energy conditions}

The weak energy condition says that the energy density of any system at
any point of spacetime for any timelike observer is positive (in the
frame of the matter this amounts to $\rho>0$ and $\rho+p\geq0$), and
when the observer moves at the speed of the light it has a well defined
limit, called the null energy condition 
($\rho+p\geq0$).  The weak and null energy conditions are the weakest of
the energy conditions (the null being even weaker than the weak), their
violation signals that the other energy conditions are also violated.
In Hawking and Ellis' book \cite{hawkingellis} the weak energy
condition is thought a physically reasonable energy condition, that 
at least all classical 
systems should obey.  Afterwards it was found that it could be violated
for quantum systems, such as in the Casimir effect and Hawking
evaporation (see \cite{klinkhammer} for a short review).  It was
further found that for quantum systems in classical gravitational
backgrounds the weak or null energy conditions could only be violated
in small amounts, and a violation at a given time through the
appearance of a negative energy state, would be overcompensated by the
appearance of a positive energy state soon after. This idea gave rise
to the averaged energy condition \cite{roman1}, and to the quantum
inequalities which, being intermediate between the pointwise energy
conditions and the averaged energy conditions, limit the magnitude of
the negative energy violations and the time for which they are allowed
to exist, yielding information on the distribution of the negative
energy density in a finite neighborhood
\cite{fordroman1,fordroman2,fordroman3}.  It seems that the situation
has changed drastically, it has been now shown that even classical systems,
such as those built from scalar fields non-minimally coupled to
gravity, violate all the energy conditions \cite{barcelovisser1} 
(see also \cite{mcinnes} for other violations of the energy 
conditions).
Thus, gradually the weak and null energy conditions, 
and with it the other energy
conditions, might be losing their status of a kind of law.
\vskip 0.4cm

\subsection{Wormhole construction: a synthesis}

Surely, this has had implications on the construction of wormholes.
First, in the original paper \cite{Morris}, Morris and Thorne
constructed wormholes by hand, that is, one gives the geometry first,
which was chosen as spherically symmetric, and then manufacture the
exotic matter accordingly. The engineering work was left to an absurdly
advanced civilization, which could manufacture such matter and construct
these wormholes.  Then, once it was understood that quantum effects
should enter in the stress-energy tensor, a self-consistent wormhole
solution of semiclassical gravity was found \cite{hochbergetal},
presumably obeying the quantum inequalities. These quantum inequalities
when applied to wormhole geometries imply that the exotic matter is
confined to an extremely thin band of size only slightly larger than
the Planck length, in principle preventing traversability
\cite{fordroman2}.  Finally with the realization that nonminimal scalar
fields violate the weak energy condition a set of self-consistent
classical wormholes was found \cite{barcelovisserPLB99}.
It is fair to say that, though outside this mainstream, classical
wormholes were found by Homer Ellis back in 1973 \cite{homerellis}, and
related self-consistent solutions were found by Kirill Bronnikov in
1973 \cite{bronikovWH}, Takeshi Kodama in 1978 \cite{kodama}, and
G\'erard Cl\'ement in 1981 \cite{clementWH}, these  papers written much
before the wormhole boom originated from Morris and Thorne's work
\cite{Morris} (see \cite{clement1} for a short account of these
previous solutions). A self-consistent Ellis wormhole was found
again by Harris \cite{harris} by solving, through an exotic scalar
field, an exercise for students posed in \cite{Morris}.

\vskip 0.4cm

\subsection{Further wormhole construction}
 
Traversable wormhole theory achieved the end of its first stage after the 
writing of the monograph on the subject by Visser in 1995 \cite{Visser}. 
This monograph is fairly complete on citations, so we refer the reader 
to it for a bibliographic search up to 1995. We refer here to some 
developments afterwards, quoting older references when appropriate.

\vskip 0.3cm
{\it Further wormhole construction in general relativity:}
\vskip 0.1cm
Visser led the way through several works.  Indeed,  Visser
\cite{visser1} constructed wormholes with polyhedrical symmetry in
1989, generalized a suggestion of Roman for a configuration with two wormholes
\cite{Morris} into a Roman ring \cite{visserRing}, he
started a study on generic dynamical traversable wormhole throats
\cite{hochvisserWHthroats} in 1997, found classically consistent
solutions with scalar fields \cite{barcelovisser1} in 1999, and has
also found self-dual solutions \cite{dadichetalSelfdual}.  Other
authors have made also interesting studies.  Before Visser's book we
can quote the paper by Frolov and Novikov, where they mix wormhole and
black hole physics \cite{frolov1}.  After the book, particularly 
interesting wormholes with
toroidal symmetry were found by Gonz\'alez-D\'{\i}as \cite{gonzalez},
wormhole solutions inside cosmic strings were found by Cl\'ement
\cite{clementWHstrings}, and Aros and Zamorano \cite{aroszamorano},
wormholes supported by strings by Schein, Aichelburg and Israel, 
\cite{shein1},  rotating wormholes were found by Teo \cite{teo}, consistent
solutions of the Einstein$-$Yang-Mills theory in connection with
primordial wormhole formation were found in \cite{nojiri}, 
theorems for the impossibility of existence of wormholes in 
some Einstein-scalar theories were discussed by Saa \cite{saa}, 
wormholes
with stress-energy tensor of massless neutrinos and other massless
fields by Krasnikov \cite{kras}, wormholes made of a crossflow of dust
null streams were discussed by Hayward \cite{haywardnulldust} and
Gergely \cite{gergely}, and self consistent charged solutions were
found by Bronnikov and Grinyok \cite{bronnikovgrinyok2}.
\vskip 0.3cm

{\it Wormhole construction with arbitrarily small violations of the
energy conditions:}
\vskip 0.1cm
One of the main areas in wormhole research is to try 
to avoid as much as possible the violation 
of the null energy condition. For static wormholes 
the null energy condition is violated \cite{Morris,Visser}. 
Several attempts have been made to overcome somehow this 
problem: Morris and Thorne already had tried 
to minimize the violating region  in the original 
paper \cite{Morris}, 
Visser \cite{visser1} found solutions where 
observers can pass the throat without interacting 
with the exotic matter, which was pushed to the corners, and 
Kuhfittig \cite{kuhfittig} has found that the region 
made of exotic matter can be made arbitrarily small. 
For dynamic wormholes, the violation of the 
weak energy condition can be avoided, but the 
null energy condition, more precisely the averaged 
null energy condition is not preserved 
\cite{hochvisserWHthroats,hochvisserPRL98,hochvisserPRD98}, 
although in \cite{visser2003} it has been found that the 
quantity of violating matter can be made arbitrarily small, 
a result in line with \cite{kuhfittig} for static wormholes. 
\vskip 0.3cm

{\it Wormhole construction with a cosmological 
constant $\Lambda$:}
\vskip 0.1cm
Some papers have added a cosmological constant 
to the wormhole construction. Kim \cite{kimwh} found thin shell 
solutions in the spirit of Visser \cite{Visser}, Roman
\cite{romanLambda} found a wormhole solution inflating in time to test
whether one could evade the violation of the energy conditions, Delgaty
and Mann \cite{delgaty} looked for new wormhole solutions with
$\Lambda$, and DeBenedectis and Das \cite{benedectis} found a general
class with a cosmological constant. Here, we will further study 
wormholes in a spacetime with a cosmological constant, as will be detailed 
below. 
\vskip 0.9cm

{\it Wormhole construction in other theories of gravitation:}
\vskip 0.1cm
In alternative theories to general relativity wormhole solutions have
been worked out. In higher dimensions solutions have been found by
Chodos and Detweiler \cite{chodos}, Cl\'ement \cite{clementhigherD},
and DeBenedictis and Das \cite{deBenedictisDas}, in Brans-Dicke theory
by Nandi and collaborators \cite{nandi}, in Kaluza-Klein theory by Shen
and collaborators \cite{shen}, in Einstein-Gauss-Bonnet by Kar
\cite{kar}, Anchordoqui and Bergliaffa found a wormhole solution in a
brane world scenario \cite{anchordoqui} further examined by Barcel\'o
and Visser \cite{barcelovisserBrane}, and Koyama, Hayward and Kim
\cite{koyamahaywardkim} examined a two-dimensional dilatonic theory.

\subsection{Stability}

To know the stability of an object against several types of
perturbation is always an important issue. Wormholes are not an
exception.  Not many works though are dedicated to the stability theory
of wormholes, although the whole formalism developed for relativistic
stars and black holes could be readily used in wormholes.  Visser
\cite{visserNPB}, Poisson and Visser \cite{poisson} and Ishak and Lake
\cite{lake} studied the stability of wormholes made of thin shells and
found, in the parameter space
$({\cal P}/\Sigma)\times$(throat$\,$radius/mass), 
where $\Sigma$ and $\cal P$ are, respectively, the surface energy
density and surface tangential pressure, those wormholes for which
there are stable solutions.  For the Ellis' drainhole
\cite{homerellis}, Armend\'ariz-Pic\'on \cite{picon} finds that it is
stable against linear perturbations, whereas Shinkai and Hayward
\cite{shinkai} find this same class unstable to nonlinear
perturbations.  Bronnikov and Grinyok
\cite{bronnikovgrinyok2,bronnikovgrinyok1} found that the consistent
wormholes of Barcel\'o and Visser \cite{barcelovisserPLB99} are
unstable.
\vskip 0.4cm

\subsection{Wormholes as time machines} 
An important side effect of wormholes is that they can be converted
into time machines, by performing a sufficient delay to the time of one
mouth in relation to the other. This can be done either by the special
relativistic twin paradox method \cite{mty} or by the general
relativistic redshift way \cite{frolovnovikovTM}.  The importance of
wormholes in the study of time machines is that they provide a
non-eternal time machine, where closed timelike curves appear to the
future of some hypersurface, the chronology horizon (a special case of a
Cauchy horizon) which is generated in a compact region in this case.  
Since time
travel to the past is in general unwelcome, it is possible to test
whether classical or semiclassical effects will destroy the time
machine. It is found that classically it can be easily stabilized
\cite{mty,Visser}.  Semiclassically, there are calculations that favor
the destruction \cite{kimthorne,hawking}, leading to 
chronology protection \cite{hawking}, 
others that maintain the
machine \cite{lyu,visserRing}.
Other simpler systems that simulate a wormhole, 
such as Misner spacetime which is a species of two-dimensional 
wormhole, have been studied more thoroughly, with 
no conclusive answer. For Misner spacetime 
the debate still goes on, favoring 
chronology protection \cite{hiskonk}, 
disfavoring it \cite{gottLI}, and back in favoring \cite{hiscock}. 
The upshot is that semiclassical calculations will not settle the 
issue of chronology protection \cite{visserCP}, one needs 
a quantum gravity, as has been foreseen 
sometime before by Thorne \cite{thorneGRG13}. 
\vskip 0.4cm

\subsection{Towards a unified view: From stars to wormholes}

There is now a growing consensus that wormholes are in the same chain
of stars and black holes. For instance, Gonz\'alez-D\'{\i}as
\cite{gonzalez} understood that an enormous pressure on the center
ultimately meant a negative energy density to open up the tunnel,
DeBenedectis and Das \cite{benedectis} mention that the stress-energy
supporting the structure consists of an anisotropic brown dwarf `star',
and the wormhole joining one Friedmann-Robertson-Walker universe with
Minkowski spacetime or joining two  Friedmann-Robertson-Walker universes
\cite{hochvisserWHthroats} could be interpreted, after further 
matchings,  as a wormhole joining a
collapsing (or expanding) star to Minkowski spacetime or a wormhole
joining two dynamical stars, respectively. It has also been recognized, 
and emphasized by Hayward \cite{hayward}, 
that wormholes and black holes can be treated in a unified way, the
black hole being described by a null outer trapped surface, and the
wormhole by a timelike outer trapped surface, this surface being the
throat where incoming null rays start to diverge
\cite{hochvisserPRD98,hayward}. Thus, it seems there is a continuum of
objects from stars to wormholes passing through black holes, where
stars are made of normal matter, black holes of vacuum, and wormholes
of exotic matter. Although not so appealing perhaps, 
wormholes could be called exotic stars. 
\vskip 0.4cm

\subsection{Astrophysical signatures}

Stars are common for everyone to see, black holes also inhabit the
universe in billions, so one might tentatively guess that wormholes,
formed or constructed from one way or another, can also appear in
large amounts. If they inhabit the cosmological space, they will
produce microlensing effects on point sources at non-cosmological
distances \cite{cramer}, as well as at cosmological distances, in this
case gamma-ray bursts could be the objects microlensed
\cite{torres,safonova1}. If peculiarly large wormholes will produce
macrolensing effects \cite{safonova2}. 
\vskip 0.4cm

\subsection{Aim of this paper}  

In this paper we  extend the Morris$\,$-Thorne wormhole solutions
\cite{Morris} by including a cosmological constant $\Lambda$.
Morris$\,$-Thorne wormholes, with $\Lambda=0$, have two asymptotically
flat regions.  By adding a positive cosmological constant, $\Lambda>0$,
the wormholes have two asymptotically de Sitter regions, and by adding
a negative cosmological constant, $\Lambda<0$, the wormholes have two
asymptotically anti-de Sitter regions.  There are a number of reasons
to study wormholes with generic $\Lambda$ that 
a technologically absurdly 
advanced civilization might construct. For $\Lambda>0$, we know
that an inflationary phase of the ultra-early universe demands it, and
moreover, from recent astronomical observations, it seems that we live
now in a world with $\Lambda>0$.  On the other hand, $\Lambda<0$ is the
vacuum state for extended theories of gravitation such as supergravity
and superstring theories, and, in addition, 
even within general relativity, a negative
cosmological constant permits solutions of black holes with horizons
with topology different from the usual spherical \cite{lemos,zanchin}
(see \cite{lemosreview} for a review), which could be turned into wormhole
solutions by adding some exotic matter, 
although we do not attempt it here.

We follow the spirit of the Morris and Thorne paper \cite{Morris}, in that
wormhole theory is a good tool to teach general relativity and a
subject that attracts students. We analyze distributions of
matter similar to \cite{Morris} but now with generic $\Lambda$, i.e, we
analyze spherically symmetric and static traversable Morris$\,$-Thorne
wormholes in the presence of a cosmological constant. 
The more complicated issue of 
the formalism of junction
conditions, that Morris and Thorne so well evaded \cite{Morris}, is
here treated also in a pedagogical way through the direct use of
Einstein field equation, and the matter content of the thin shell
separating the wormhole from the exterior spacetime is found.  In this
way, an equation connecting the radial tension at the mouth with the
tangential surface pressure of the thin shell is derived. The structure
as well as several physical properties and characteristics of
traversable wormholes due to the effects of the cosmological term are
studied.  We find that for asymptotically flat and anti-de Sitter
spacetimes the surface tangential pressure $\cal P$ of the thin shell
is always strictly positive, whereas for de Sitter spacetime it can
take either sign as one could expect, being negative (tension) for
relatively high $\Lambda$ and high wormhole radius, positive for
relatively high mass and small wormhole radius, and zero in-between.
Finally, some specific solutions with $\Lambda$, based on the
Morris$\,$-Thorne solutions, are provided. In presenting these solutions
we dwell mostly on the case $\Lambda=0$, and $\Lambda>0$, and comment
briefly on $\Lambda<0$.  The plan of the paper is as follows: In
section 2 we present the Einstein field equation for a wormhole metric
and perform the junction to an external asymptotically Minkowski, 
de Sitter, or 
anti-de Sitter spacetime. In section 3 we give some wormhole
geometries, analogous to \cite{Morris} having $\Lambda=0$, $\Lambda>0$
and $\Lambda<0$, and study some of their properties.  In section 4 we
conclude.

\section{Einstein field equation for wormholes with a generic cosmological 
constant $\Lambda$} 
 
\subsection{The Einstein field equation with generic $\Lambda$, setting the 
nomenclature}

To set the nomenclature, the Einstein field equation with a
cosmological constant is given, in a coordinate basis, by
G$_{\mu\nu}+\Lambda g_{\mu\nu}=8\pi G c^{-4} T_{\mu\nu}$, in which
$G_{\mu\nu}$ is the Einstein tensor, given by
$G_{\mu\nu}=R_{\mu\nu}-\frac{1}{2}g_{\mu\nu}R$, $R_{\mu\nu}$ is the
Ricci tensor, which is defined as a contraction of the Riemann (or
curvature) tensor, $R_{\mu\nu}=R^{\alpha}_{\;\; \mu \alpha\nu}$, and
$R$ is the scalar curvature defined as a contraction of the Ricci
tensor, $R=R_{\alpha}^{\alpha}$.  The Riemann tensor is a function of
the second order derivatives of the metric components $g_{\mu\nu}$.
$T_{\mu\nu}$ is the stress-energy tensor of the matter, and $\Lambda$
the cosmological constant \cite{MTW}.
 
\subsubsection{The spacetime metric} 
 
We will be interested in the spacetime metric, representing a 
spherically symmetric and static wormhole, given by \cite{Morris} 
\begin{equation} 
ds^2=-e ^{2\Phi(r)} \,c^2\,dt^2+\frac{dr^2}{1- b(r)/r}+r^2 
\,(d\theta ^2+\sin ^2{\theta} \, d\phi ^2) \,,\label{metricwormhole} 
\end{equation} 
where $(c\,t,r,\theta,\phi)$ are the usual spacetime spherical 
coordinates, and 
$\Phi(r)$ and $b(r)$ are arbitrary functions of the radial 
coordinate $r$. $\Phi(r)$ is designated the redshift function, 
for it is related to the gravitational redshift, and $b(r)$ is 
denominated the shape function, because as can be shown by 
embedding diagrams, it determines the shape of the wormhole 
\cite{Morris}. The radial coordinate has a range that increases 
from a minimum value at $r_{\rm o}$, corresponding to the wormhole 
throat, to a maximum $a$ corresponding to the mouth. At $r_{\rm o}$ 
one has to join smoothly this spherical volume 
to another spherical volume copy with $r$ ranging again from 
$r_{\rm o}$ to $a$, see Figure 1. In addition, one has then to join 
each copy to the external spacetime from $a$ to $\infty$,  
as will be done. 
\vskip 0.5cm
\centerline{\epsffile{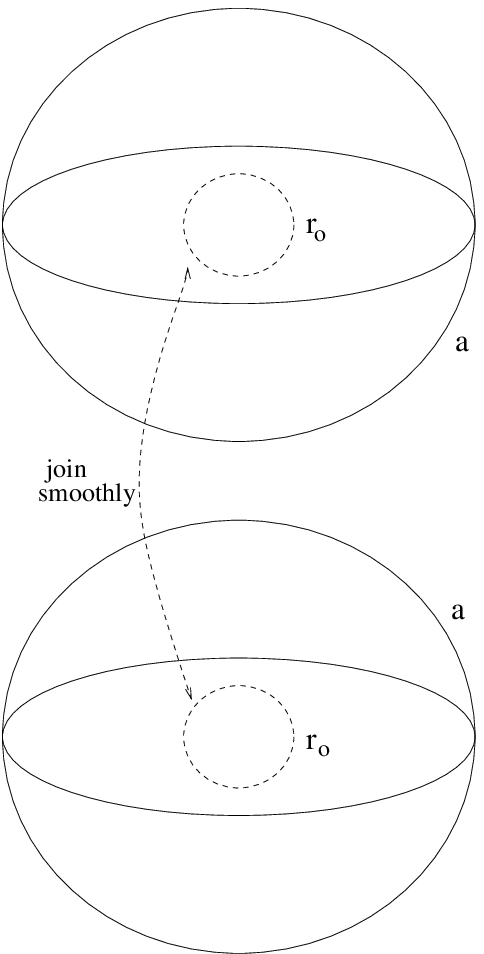}}
{\noindent {\small Figure 1 - The two copies of the 
spherical volume from $r_{\rm o}$ to $a$ have to be 
smoothly joined at $r_{\rm o}$.}}
\vskip 0.5cm

The mathematical analysis and the physical interpretation will be 
simplified using a set of orthonormal basis vectors. 
These may be interpreted as the proper reference frame of a set of 
observers who remain at rest in the coordinate system 
$(ct,r,\theta,\phi)$, with $(r,\theta,\phi)$ fixed. 
Denote the basis vectors in the coordinate system as 
${\bf e}_{t}$, ${\bf e}_{r}$, ${\bf e}_{\theta}$, 
and ${\bf e}_{\phi}$.
Then, using the 
following transformation, ${\bf e}_{\hat{\mu}}=\Lambda 
^{\nu}_{\; \hat{\mu}}\; {\bf e}_{\nu}$,  
with 
\begin{equation} 
\Lambda ^{\nu}_{\; \hat{\mu}}={\rm diag} \left 
[e^{-\Phi},(1-b/r)^{1/2},r^{-1},(r \sin \theta)^{-1} \right ]\,,
\end{equation} 
where the notation means that the non-diagonal terms of the 
matrix are zero, one finds 
\begin{eqnarray} 
\left \{ \begin{array}{l} 
{\bf e}_{\hat{t}}=e^{-\Phi} \;{\bf e}_{t}\\ 
{\bf e}_{\hat{r}}=(1-b/r)^{1/2}\; {\bf e}_{r}\\ 
{\bf e}_{\hat{\theta}}=r^{-1} \;{\bf e}_{\theta}\\ 
{\bf e}_{\hat{\phi}}=(r \sin \theta)^{-1}\; {\bf e}_{\phi}\,.
              \end{array} \right. 
\end{eqnarray} 
In this basis the metric components assume their Minkowskian 
form, given by, 
\begin{equation} 
g_{\hat{\mu}\hat{\nu}}=\eta _{\hat{\alpha}\hat{\beta}}={\rm 
diag} (-1,1,1,1). 
\end{equation} 
In the orthonormal reference frame, the Einstein field equation
with a generic 
cosmological constant, is given by 
\begin{equation} 
G_{\hat{\mu}\hat{\nu}}+\Lambda \eta_{\hat{\mu}\hat{\nu}}= 
\frac{8\pi G}{c^{4}}T_{\hat{\mu}\hat{\nu}}\,.
\label{einsteinorto} 
\end{equation} 
The Einstein tensor, given  in the orthonormal reference frame by 
$G_{\hat{\mu}\hat{\nu}}=R_{\hat{\mu}\hat{\nu}}-\frac{1}{2}R\, 
g_{\hat{\mu}\hat{\nu}}$, yields for the metric (\ref{metricwormhole})
the following non-zero components 
\begin{eqnarray} 
G_{\hat{t}\hat{t}}&=&\frac{b'}{r^2} \label{Einsteintt}\,, \\ 
G_{\hat{r}\hat{r}}&=&-\frac{b}{r^3}+ 2 \left(1-\frac{b}{r} 
\right) \frac{\Phi'}{r}  \label{Einsteinrr}\,,  \\ 
G_{\hat{\theta}\hat{\theta}}&=&\left(1-\frac{b}{r} 
\right)\left[\Phi ''+ (\Phi')^2- \frac{b'r-b}{2r(r-b)}\Phi'- 
\frac{b'r-b}{2r^2(r-b)}+\frac{\Phi'}{r} \right]\,, 
\label{Einsteintheta}\\
G_{\hat{\phi}\hat{\phi}}&=&G_{\hat{\theta}\hat{\theta}}\,,
\label{Einsteinphi}
\end{eqnarray} 
where a prime denotes a derivative with respect to the radial 
coordinate $r$. 
 
\subsubsection{The stress-energy tensor} 
 
The Einstein field equation requires that the Einstein tensor be 
proportional to the stress-energy tensor. In the orthonormal basis 
the stress-energy tensor, $T_{\hat{\mu}\hat{\nu}}$, must have an 
identical algebraic structure as the Einstein tensor components, 
$G_{\hat{\mu}\hat{\nu}}$, i.e., Equations 
(\ref{Einsteintt})-(\ref{Einsteinphi}). Therefore, the only 
non-zero components of $T_{\hat{\mu}\hat{\nu}}$ are  
$T_{\hat{t}\hat{t}}$, $T_{\hat{r}\hat{r}}$,  
$T_{\hat{\theta}\hat{\theta}}$, and $T_{\hat{\phi}\hat{\phi}}$. 
These are 
given an immediate physical interpretation, 
\begin{eqnarray} 
T_{\hat{t}\hat{t}}&=&\hskip 1mm\rho(r)c^2  \label{rhoWH}\,, \\ 
T_{\hat{r}\hat{r}}&=&\hskip -3mm -\tau (r)  \label{tauWH}\,, \\ 
T_{\hat{\theta}\hat{\theta}}&=&\hskip 1mmp(r) \label{pressureWH}, \\ 
T_{\hat{\phi}\hat{\phi}}&=&\hskip 1mmp(r) \label{pressureWH2}\,, 
\end{eqnarray} 
in which $\rho(r)$ is the energy density, $\tau (r)$ is the radial 
tension, with $\tau (r)=-p_r(r)$, i.e., it is the negative of the 
radial pressure, $p(r)$ is the pressure measured in the tangential  
directions, orthogonal to the radial direction. 
$T_{\hat{\mu}\hat{\nu}}$ may include surface quantities
as we will see. 
 
\subsubsection{The cosmological constant 
and the total stress-energy tensor} 
 
To obtain a physical interpretation of the cosmological constant, 
one may write the Einstein field equation in 
the following manner: $G_{\hat\mu\hat\nu}=8\pi G 
c^{-4}(T_{\hat\mu\hat\nu}+T_{\hat\mu\hat\nu}^{({\rm vac})})$, in which 
$T_{\hat\mu\hat\nu}^{({\rm vac})}=-g_{\hat\mu\hat\nu}
(\Lambda c^4/(8\pi G))$ may 
be interpreted as the stress-energy tensor associated with the 
vacuum, and in the orthonormal reference frame is given by 
\begin{equation} 
T^{({\rm vac})}_{\hat{\mu}\hat{\nu}}={\rm diag} \left[\Lambda 
c^4/(8\pi G),-\Lambda c^4/(8\pi G),-\Lambda c^4/(8\pi G),-\Lambda 
c^4/(8\pi G)\right]. 
\end{equation} 
We see it is thus 
possible to adopt the viewpoint that the cosmological term 
is an integral part of the stress-energy tensor, being
considered as a fluid.  
Accordingly, we can define the total stress-energy tensor, 
$\overline{T}_{\hat{\mu}\hat{\nu}}$, as 
\begin{equation} 
\overline{T}_{\hat{\mu}\hat{\nu}}=T_{\hat{\mu}\hat{\nu}}+T^{({\rm 
vac})}_{\hat{\mu}\hat{\nu}} 
\end{equation} 
such that $G_{\hat{\mu}\hat{\nu}}=8\pi G 
c^{-4}\overline{T}_{\hat{\mu}\hat{\nu}}$. Thus, the components of 
the total stress-energy tensor of the wormhole, $\bar{\rho}(r)$, 
$\bar{\tau}(r)$ and $\bar{p}(r)$, are given by the following 
functions 
\begin{eqnarray} 
\bar{\rho}(r)&=& \rho(r) +\frac{c^2}{8\pi G}\,\Lambda   
\label{totalrho}\,, \\ 
\bar{\tau}(r)&=& \tau(r) +\frac{c^4}{8\pi G}\,\Lambda   
\label{totaltau}\,, \\ 
\bar{p}(r)&=& p(r) -\frac{c^4}{8\pi G}\,\Lambda 
\label{totalp}\,. 
\end{eqnarray} 
This viewpoint may be interesting to adopt in some cases.

\subsubsection{The Einstein equations} 
 
We are interested in matching the interior solution, 
whose metric is given by 
Equation (\ref{metricwormhole}), to an exterior vacuum solution, 
which will be considered below. Using  Equation 
(\ref{einsteinorto}) and equating Equations  
(\ref{Einsteintt})-(\ref{Einsteintheta}) 
with (\ref{rhoWH})-(\ref{pressureWH})  we obtain the 
following set of equations
\begin{eqnarray} 
&\rho(r)=\frac{c^2}{8\pi G} \left(\frac{b'}{r^2}-\Lambda \right)  
\label{rhoWHlambda}\,,\\ 
&\tau (r)=\frac{c^4}{8\pi G} \left[\frac{b}{r^3}-2 
\left(1-\frac{b}{r} 
\right) \frac{\Phi'}{r}- \Lambda \right]  \label{tauWHlambda}\,,\\ 
& p(r)=\frac{c^4}{8\pi G} \left 
\{\left(1-\frac{b}{r}\right)\left[\Phi ''+ (\Phi')^2- 
\frac{b'r-b}{2r^2(1-b/r)}\Phi'- 
\frac{b'r-b}{2r^3(1-b/r)}+\frac{\Phi'}{r} \right] + \Lambda \right 
\}   \label{pressureWHlambda}\,. 
\end{eqnarray} 
 
By taking the derivative of Equation (\ref{tauWHlambda}) 
with respect to the radial coordinate  
$r$,  and eliminating $b'$ and 
$\Phi''$, given in Equation (\ref{rhoWHlambda}) and Equation 
(\ref{pressureWHlambda}), respectively, we obtain the following 
equation 
\begin{equation} 
\tau '=(\rho c^2-\tau )\Phi '-\frac{2}{r}(p+\tau ) 
\label{tauderivative} \,. 
\end{equation} 
Equation (\ref{tauderivative}) is 
the relativistic Euler equation, or the hydrostatic 
equation for equilibrium for the material threading the wormhole, 
and can also be obtained using the 
conservation of the stress-energy tensor, 
$T^{\hat{\mu}\hat{\nu}}_{\;\;\;\;;\,\hat{\nu}}=0$, putting 
$\hat{\mu}=r$. The conservation of the stress-energy tensor, in 
turn can be deduced from the Bianchi identities, which are 
equivalent to $G^{\hat\mu \hat\nu}_{\;\;\;\;;\,\hat\nu}=0$. 
 
\subsubsection{Method for solving the 
Einstein equations} 
 
The conventional approach to solving the  Einstein equations 
would be to assume a specific and plausible type of matter or 
fields for the source of the stress-energy tensor. One would then 
derive equations of state for the radial tension and the tangential  
pressure, as functions of the energy density. These equations of 
state, together with the three field equations would provide  
the geometry of the spacetime given in terms of the metric, 
$g_{\mu\nu}$, as we would have five equations for five unknown 
functions, i.e., $b(r)$, $\Phi(r)$, $\rho$, $\tau$ and $p$. 
Morris and Thorne's approach \cite{Morris}, which will be followed 
in this paper,  
differs as they first fixed 
a convenient geometry for a wormhole solution and then derived the 
matter distribution for the respective solution 
(see \cite{benedectis} for a careful analysis of the various approaches).

\subsection{Construction of a wormhole with generic $\Lambda$. I:
General comments}

\subsubsection{The mathematics of embedding} 
 
We can use embedding diagrams to represent a wormhole and extract 
some useful information for the choice of the shape function,  
$b(r)$, which will be used in the specific solutions considered 
below. Due to the spherically symmetric nature of the problem, one 
may consider an equatorial slice, $\theta=\pi/2$, without loss of 
generality. The respective line element, considering a fixed 
moment of time, $t={\rm const}$, is given by 
\begin{equation} 
ds^2=\frac{dr^2}{1- b(r)/r}+r^2 \, d\phi ^2\,. \label{surface1}
\end{equation} 
To visualize this slice, one embeds this metric into 
three-dimensional Euclidean space, whose metric 
can be written in 
cylindrical coordinates, $(r,\phi,z)$, as
\begin{equation} 
ds^2=dz^2+dr^2+r^2 \, d\phi ^2 \,.
\end{equation} 
Now,  in the three-dimensional Euclidean 
space the embedded surface has equation $z=z(r)$, and 
thus the metric of the surface can be written as, 
\begin{equation} 
ds^2=\left [1+\left( \frac{dz}{dr}\right)^2\right] dr^2+r^2 \, 
d\phi ^2 \,. \label{surface2}
\end{equation} 
Comparing Equation (\ref{surface2}) 
with  (\ref{surface1}) 
we have the 
equation for the embedding surface, given by 
\begin{equation} 
\frac{dz}{dr}=\pm \left(\frac{r}{b(r)}-1 \right)^{-1/2} 
\label{lift}\,. 
\end{equation} 
To be a solution of a wormhole, the geometry has a minimum radius, 
$r=b(r)=r_{\rm o}$, denoted as the throat, at which the embedded surface 
is vertical, i.e., $dz/dr \rightarrow \infty$, see Figure 2. Outside 
the wormhole, far from the 
mouth, space can be asymptotically flat, de Sitter, or 
anti-de Sitter.
\vskip 0.5cm
\centerline{\epsffile{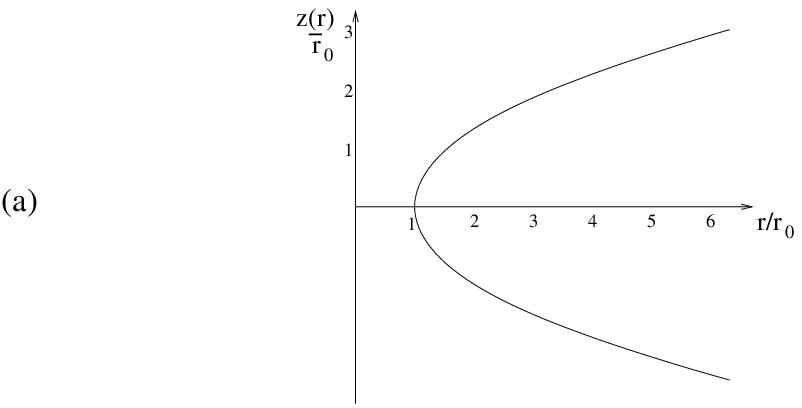  }}
\vskip 5mm
\centerline{\epsffile{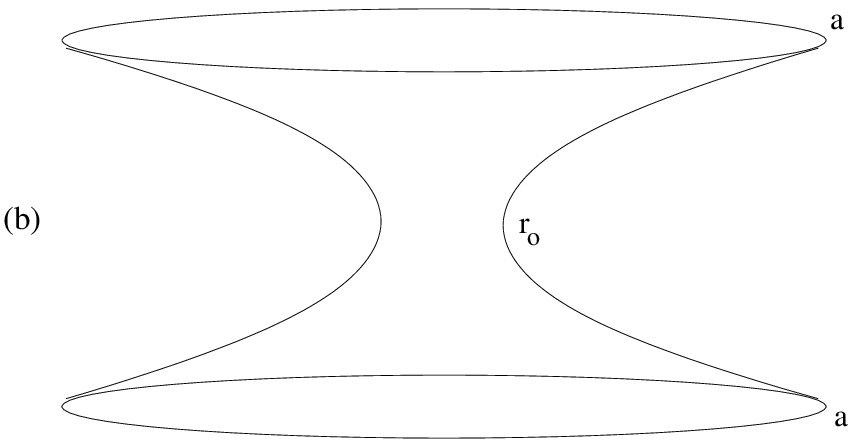}}
\vskip 1mm
{\noindent {\small Figure 2 - (a) 
The embedding diagram of a two-dimensional 
section ($t=\,$constant, $\theta=\pi/2$) of the wormhole 
in three-dimensional Euclidean space, 
here $a/r_{\rm o}=6$; (b) For the full visualization 
of the surface sweep through a $2\pi$ rotation around the 
$z-$axis. Figure 1 with one dimension less is equivalent to 
this figure.}}
\vskip 0.5cm
One can define the proper radial distance for the upper part of 
the wormhole $z>0$ as 
\begin{equation} 
l(r)= \int ^{r}_{r_{\rm o}} \,\frac{dr}{\left[1-b(r)/r \right]^{1/2}} 
\label{properradialcoordinate} \,, 
\end{equation}
and for the lower part  $z<0$ as 
\begin{equation} 
l(r)= -\int ^{r}_{r_{\rm o}} \,\frac{dr}{\left[1-b(r)/r \right]^{1/2}} 
\label{properradialcoordinatelower} \,.
\end{equation} 
The maximum upper limit of integration is $a$, the radius of the 
wormhole mouth. 
The shape function $b(r)$ should be positive and such that 
$b/r<1$ 
in order to have  $\sqrt{r/b-1}$  real. For generic cosmological constant 
this may not be possible. For instance, in vacuum for sufficiently 
large radii Equation (\ref{lift}) becomes imaginary, in which case 
the embedding process is no longer valid. However, the importance 
of the embedding is near the throat where a special condition, the 
flare out condition, should be obeyed.

\subsubsection{Exotic matter} 
 
Following \cite{Morris} closely we will see that 
the wormhole needs exotic matter, i.e, matter that does not 
obey the null energy condition, and thus does not obey the 
weak or any other energy condition. The null energy 
condition applied to the matter considered in 
(\ref{totalrho})-(\ref{totaltau}) is $\bar\rho\,c^2-\bar\tau>0$. 
Thus a good way to define exoticity is through the parameter 
$\xi$ defined as \cite{Morris} 
$\xi=\frac{\bar{\tau}-\bar{\rho}c^2}{|\bar{\rho}c^2|}$.
This parameter $\xi$ is dimensionless, and when positive signals 
exotic matter. 
Using equations (\ref{rhoWHlambda})-(\ref{tauWHlambda}) one finds 
\begin{equation} 
\xi 
=\frac{\bar{\tau}-\bar{\rho}c^2}{|\bar{\rho}c^2|}=\frac{b/r-b'- 
2r(1-b/r)\Phi'}{|b'-\Lambda r^2|}   \label{exoticity}\,. 
\end{equation}
To be a solution of 
a wormhole, one needs to impose that the throat flares out, 
as in Figure 2. 
Mathematically, this flaring-out condition entails that the 
inverse of the embedding function $r(z)$, must satisfy 
$d^2r/dz^2>0$ near the throat $r_{\rm o}$. Differentiating 
$dr/dz=\pm (r/b(r)-1)^{1/2}$ with respect to $z$, we have 
\begin{equation} 
\frac{d^2r}{dz^2}=\frac{b-b'r}{2b^2}>0   \label{flareout}\,. 
\end{equation} 
Combining Equation (\ref{exoticity}) with Equation 
(\ref{flareout}), the exoticity function takes the form 
\begin{equation} 
\xi =\frac{2b^2}{r|b'-\Lambda r^2|}\;\frac{d^2r}{dz^2}-2r 
\left(1-\frac{b}{r}\right)\frac{\Phi'}{|b'-\Lambda r^2|}  \,. 
\end{equation} 
Considering the finite character of $\rho$, and therefore of $b'$, 
and the fact that $(1-b/r)\Phi' \rightarrow 0$ at the throat, we 
have the following relationship 
\begin{equation} 
\xi(r_{\rm o}) =\frac{\bar \tau_0-\bar \rho_0c^2}{|\bar\rho_0c^2|}>0 \,. 
\end{equation} 
Thus matter at the throat is exotic
(see \cite{Morris,Visser,hochvisserWHthroats} 
for a detailed discussion).

\subsection{Construction of a wormhole with generic $\Lambda$.  
II: Interior and exterior solutions, and junction conditions} 
 
We will distinguish the interior cosmological 
constant, $\Lambda_{\rm int}$, from the exterior 
one, $\Lambda_{\rm ext}$. 
Equations (\ref{rhoWHlambda})-(\ref{pressureWHlambda}) demonstrate 
that once the geometry is fixed by the redshift function 
$\Phi(r)$, and the shape function $b(r)$, the inclusion of the 
cosmological constant $\Lambda_{\rm int}$ 
will shift the respective values of 
$\rho(r)$, $\tau (r)$ and $p(r)$ and might help in minimizing 
the violation of the energy conditions.

\subsubsection{Interior solution of the Einstein equations 
with generic $\Lambda_{\rm int}$}
 
To find an interior solution of the Einstein equations 
with generic $\Lambda$, we 
combine the analysis developed in the previous sections, taking into 
account the notation that $\Lambda_{\rm int}$ represents the 
cosmological constant associated with the interior solution. The 
respective Einstein equations provide the following relationships 
\begin{eqnarray} 
&\rho(r)=\frac{c^2}{8\pi G} \left(\frac{b'}{r^2}-\Lambda_{\rm int} \right) 
\label{rhoWHlambdaint}\,,\\ 
&\tau (r)=\frac{c^4}{8\pi G} \left[\frac{b}{r^3}-2 
\left(1-\frac{b}{r} 
\right) \frac{\Phi'}{r}- \Lambda_{\rm int} \right] 
\label{tauWHlambdaint}\,,\\ 
&\hskip -0.7cm p(r)=\frac{c^4}{8\pi G} \left 
\{\left(1-\frac{b}{r}\right)\left[\Phi ''+ (\Phi')^2- 
\frac{b'r-b}{2r^2(1-b/r)}\Phi'- 
\frac{b'r-b}{2r^3(1-b/r)}+\frac{\Phi'}{r} \right] + \Lambda_{\rm 
int} \right \}    \label{pressureWHlambdaint}. 
\end{eqnarray} 
The metric quantities should carry a subscript ${}_{\rm int}$, but 
we do not put it in order to not overload the notation. 
It is of interest to find an expression 
for the radial tension at the throat. 
From equation (\ref{tauWHlambdaint}) one finds that at the throat 
($b(r_{\rm o})=r_{\rm o}$) the tension is 
\begin{equation} 
{\tau}(r_{\rm o})=\frac{c^4}{8\pi G 
}\left(\frac{1}{r_{\rm o}^2}-\Lambda_{\rm int} \right) \,.
\end{equation} 
Thus the radial tension at the throat is positive  
for wormholes whose structure yields 
$\Lambda_{\rm int}<\frac{1}{r_{\rm o}^2}$, 
this includes wormholes 
with negative and zero cosmological constant.
The radial tension
is negative, i.e., it is a pressure,  
for wormholes 
with the cosmological constant obeying 
$\Lambda_{\rm int}>\frac{1}{r_{\rm o}^2}$. The 
total radial tension, ${\bar \tau}(r_{\rm o})=
\tau(r_{\rm o})+ \frac{c^4}{8 \pi G} \Lambda_{\rm int}$,  is 
always positive, of course.

\subsubsection{Exterior vacuum solution of the Einstein 
equations with generic $\Lambda_{\rm ext}$} 
 
The spacetime geometry for a vacuum exterior region is simply 
determined considering a null stress-energy tensor, 
$T_{\hat{\mu}\hat{\nu}}=0$, i.e., $\rho(r)=\tau (r)=p(r)=0$. Note 
that $\Lambda_{\rm ext}$ represents the cosmological constant 
associated with the exterior solution. In the most general case, the 
exterior radial coordinate $\bar r$, 
should be different from the interior one $r$. Here we put 
them equal, both are denoted by $r$, since it simplifies the junction 
and it gives interestingly enough results.
The Einstein equations then 
reduce to 
\begin{eqnarray} 
&0=\frac{b'}{r^2}-\Lambda_{\rm ext} \label{rhovacuum}\,,\\ 
&0=\frac{b}{r^3}-2 \left(1-\frac{b}{r} \right) 
\frac{\Phi'}{r}- \Lambda_{\rm ext} \label{tauvacuum}\,,\\ 
&0= \left(1-\frac{b}{r}\right )\left[\Phi''+ 
(\Phi')^2-\frac{b'r-b}{2r^2(1-b/r)}\Phi'-\frac{b'r-b}{2r^3(1-b/r)}+ 
\frac{\Phi'}{r}\right ]+\Lambda_{\rm ext} 
\label{pressurevacuum}\,. 
\end{eqnarray} 
The metric quantities should carry a subscript ${}_{\rm ext}$, 
but again 
we have not put it as to not overload the notation. 
Solving the system of differential equations of Equations 
(\ref{rhovacuum})-(\ref{pressurevacuum}), the exterior vacuum 
solution with a cosmological constant is given by 
\begin{eqnarray} 
&ds^2=-\left(1-\frac{2GM}{c^2r}-\frac{\Lambda_{\rm ext}}{3}r^2 
\right) \,c^2\,dt^2+\frac{dr^2}{\left(1-\frac{2GM}{c^2r}-\frac{\Lambda_{\rm 
ext}}{3}r^2 \right)} \nonumber
&\\ 
&+r^2\,(d\theta ^2+\sin ^2{\theta}\, d\phi ^2)
\label{metricvacuumlambda}\,. 
\end{eqnarray} 
This metric is the unique solution to the vacuum Einstein equations for
a static and spherically symmetric spacetime with a generic 
cosmological constant. The denomination given to it depends on the sign
of $\Lambda_{\rm ext}$.  The Schwarzschild solution, which
is a particular case, is obtained by setting $\Lambda_{\rm ext} =0$.
In the presence of a positive cosmological
constant, $\Lambda_{\rm ext}>0$, the solution is designated by the
Schwarzschild-de Sitter metric. For $\Lambda_{\rm ext}<0$, we have the
Schwarzschild-anti de Sitter metric. 
For $\Lambda\neq0$, note that this metric is not asymptotically flat as
$r\rightarrow \infty$, it is either asymptotically de Sitter
($\Lambda_{\rm ext}>0$), or asymptotically anti-de Sitter
($\Lambda_{\rm ext}<0$). However, if $\Lambda_{\rm ext}$ is extremely
small, there is a range of the radial coordinate, i.e.,
$1/\sqrt{\Lambda_{\rm ext}}\gg r \gg GM/c^2$, for which the metric is
nearly flat. For values of $r$ below this range, the effect of the mass
$M$ dominates, whereas for values above this range, the effect of the
cosmological term dominates, as for very large values of $r$ the
large-scale curvature of the spacetime must be taken into account.

\bigskip 
{\it (i) The Schwarzschild spacetime, $\Lambda_{\rm ext}=0$} 
\medskip 

Equation (\ref{metricvacuumlambda}) with $\Lambda_{\rm ext}=0$ 
is the Schwarzschild solution. The full vacuum solution represents 
a black hole in a asymptotically flat spacetime.  
The factor  
$f(r)=(1-\frac{2GM}{c^2r})$ is zero 
at 
\begin{equation}
r_b=\frac{2\,G\,M}{c^2}\,,
\label{shcwarzschildhorizon}
\end{equation}
the black hole event horizon. 
Since 
the wormhole matter will fill the region up to a radius $a$ larger 
than $r_b$ this radius does not enter into the problem. 
It is important to have it in mind, since if after construction 
one finds that $r_b>a$ than the object constructed is a black hole 
rather than a wormhole.

\bigskip 
{\it (ii) The Schwarzschild-de Sitter spacetime, $\Lambda_{\rm 
ext}>0$} 
\medskip 
 
Equation (\ref{metricvacuumlambda}) with $\Lambda_{\rm ext}>0$ 
represents a black hole in asymptotically de Sitter space. If 
$0<9\Lambda_{\rm ext}(GMc^{-2})^2<1$, the factor 
$f(r)=(1-\frac{2GM}{c^2r}-\frac{\Lambda_{\rm ext}}{3}r^2)$ is zero 
at two positive values of $r$, corresponding to two real positive 
roots. Defining 
\begin{eqnarray} 
A&=&\left(\frac{3c^4}{8\Lambda_{\rm ext}G^2M^2}\right)^{1/3} 
\label{firstfactor}\; 
\sqrt[3]{-1+\sqrt{1-\frac{c^4}{9\Lambda_{\rm ext}G^2M^2}}} \,, \\ 
B&=&\left(\frac{3c^4}{8\Lambda_{\rm ext}G^2M^2}\right)^{1/3}\; 
\sqrt[3]{-1-\sqrt{1-\frac{c^4}{9\Lambda_{\rm ext}G^2M^2}}} 
\label{secondfactor}\,, 
\end{eqnarray} 
the solutions are given by 
\begin{eqnarray} 
r_b&=&\frac{2GM}{c^2}\left(-\frac{A+B}{2}-\frac{A-B}{2}\sqrt{-3} 
\right)\,, \\ 
r_c&=&\frac{2GM}{c^2}(A+B) \,. 
\end{eqnarray} 
When $\Lambda_{\rm ext}(GM/c^2)^2 \ll 1$ (see appendix A for 
details), one gets 
\begin{eqnarray} 
r_b&=&\frac{2GM}{c^2} \left[1+\frac43
\Lambda_{\rm ext}\left(\frac{G\,M}{c^2}\right)^2\right]\,, \\ 
r_c&=&\sqrt{\frac{3}{\Lambda_{\rm ext}}} 
\left(1-\frac{GM}{c^2}\sqrt{\frac{\Lambda_{\rm ext}}{3}}\right)\,. 
\end{eqnarray} 
The smaller of the values, denoted by $r=r_b$, can be considered 
as the event horizon of the vacuum black hole solution, but since 
the wormhole matter will fill the region up to a radius $a$ superior 
than $r_b$ this radius does not enter into the problem. 
The larger value, denoted 
by $r=r_c$, can be regarded as the position of the cosmological 
event horizon of the de Sitter spacetime. 
Keeping $\Lambda_{\rm ext}$ constant, but 
increasing $M$, $r=r_b$ will increase and $r=r_c$ will 
decrease. If $9\Lambda_{\rm ext}(GMc^{-2})^2=1$, both horizons 
coincide and are situated at $r=r_b =r_c=3GM/c^2$. 
Thus we will consider $9\Lambda_{\rm ext}(GMc^{-2})^2<1$. 
Particular cases are,
$\Lambda_{\rm ext} =0$ yielding the Schwarzschild solution, 
and $M=0$ yielding the de Sitter solution. 
When $r\rightarrow \infty$ the metric tends to the 
de Sitter spacetime
\begin{equation} 
ds^2=-\left(1-\frac{\Lambda_{\rm ext}}{3}r^2 \right) 
\,c^2\,dt^2+\frac{dr^2}{\left(1-\frac{\Lambda_{\rm ext}}{3}r^2 
\right)}+r^2\,(d\theta ^2+\sin ^2{\theta}\, d\phi ^2) 
\label{schwsittermassnull}\,. 
\end{equation} 
For $\Lambda_{\rm ext} \rightarrow 0$, the de Sitter metric tends to 
the Minkowskian spacetime. In the coordinates adopted above, the 
metric of the de Sitter spacetime will be singular if 
$r=(3/\Lambda_{\rm ext})^{1/2}$, but this is a mere coordinate 
singularity signaling the presence of a cosmological event 
horizon.

\bigskip 
{\it (iii) The Schwarzschild-anti de Sitter spacetime, 
$\Lambda_{\rm ext}<0$} 
\medskip 
 
For the Schwarzschild-anti de Sitter metric, with $\Lambda_{\rm 
ext}<0$, the equation $f(r)=(1-\frac{2GM}{c^2r}+\frac{|\Lambda_{\rm 
ext}|}{3}r^2)=0$ will have only one real root, therefore implying 
the existence of one horizon. If one defines
\begin{eqnarray} 
A&=&\left(\frac{3c^4}{8|\Lambda_{\rm ext}|G^2M^2}\right)^{1/3} 
\label{firstfactorads}\; 
\sqrt[3]{1+\sqrt{1+\frac{c^4}{9|\Lambda_{\rm ext}|G^2M^2}}} \,, \\ 
B&=&\left(\frac{3c^4}{8|\Lambda_{\rm ext}|G^2M^2}\right)^{1/3}\; 
\sqrt[3]{1-\sqrt{1+\frac{c^4}{9|\Lambda_{\rm ext}|G^2M^2}}} 
\label{secondfactorads}\,, 
\end{eqnarray} 
then the solution is (see appendix for details),
\begin{equation} 
r_b=\frac{2GM}{c^2}\left(A+B\right)\,.
\end{equation} 
For $|\Lambda_{\rm 
ext}|(GM/c^2)^2 \ll 1$ one obtains 
\begin{equation} 
r_b=\frac{2GM}{c^2} \left[1-\frac43
|\Lambda_{\rm ext}|\left(\frac{G\,M}{c^2}\right)^2\right]\,. 
\end{equation} 
Once again this event horizon is avoided by filling the space 
with exotic matter from the throat at $r_{\rm o}$ up to the mouth 
at $a$, where $a>r_b$ in order that the wormhole is not a 
black hole. 
If $\Lambda_{\rm ext}=0$, the metric is reduced to the 
Schwarzschild solution. If $r\rightarrow \infty$ the metric tends 
to the anti-de Sitter solution
\begin{equation} 
ds^2=-\left(1+\frac{|\Lambda_{\rm ext}|}{3}r^2 \right) 
\,c^2\,dt^2+\frac{dr^2}{\left(1+\frac{|\Lambda_{\rm ext}|}{3}r^2 
\right)}+r^2\,(d\theta ^2+\sin ^2{\theta}\, d\phi ^2) \,. 
\label{schwantisittermassnull} 
\end{equation} 
For $|\Lambda_{\rm ext}| \rightarrow 0$, the anti-de 
Sitter metric tends to the Minkowskian spacetime.

\subsubsection{Junction conditions in wormholes with generic 
$\Lambda_{\rm ext}$}

To match the interior to the exterior one needs to apply 
the junction conditions that follow from the theory of 
general relativity. 
One of the conditions imposes the continuity of the metric 
components, $g_{\mu\nu}$, across a surface, $S$, i.e., 
$g_{\mu\nu ({\rm int})}|_{S}=g_{\mu\nu ({\rm 
ext})}|_{S}\,$. This condition is not sufficient to join 
different spacetimes. One formalism of matching, that leads 
to no errors in the calculation, uses the 
extrinsic curvature of $S$ 
(or second fundamental form of the surface $S$, 
the first fundamental form being the metric on $S$) 
see, e.g, \cite{Israel}.  However, for 
spacetimes with a good deal of symmetry, such as 
spherical symmetry, one can use directly the field 
equations to make the match, see e.g., \cite{Papahamoui} 
(see also Taub \cite{Taub}). We follow this 
latter approach. Indeed, due to the high symmetries of the 
solution, we can use the Einstein equations, Equations 
(\ref{rhoWHlambda})-(\ref{pressureWHlambda}), to determine the 
energy density and stresses 
of the surface $S$ necessary to have a match between 
the interior and exterior. 
If there are no surface 
stress-energy terms at the surface $S$, 
the junction is called a boundary surface. If, on 
the other hand, surface stress-energy terms are present, the 
junction is called a thin shell.

\bigskip 
{\it (i) Matching of the metric} 
\medskip

As was mentioned above the unique vacuum static and spherically 
symmetric solution, in the presence of a non-vanishing 
cosmological constant, is given by Equation 
(\ref{metricvacuumlambda}). A wormhole with finite dimensions, in 
which the matter distribution extends from the throat, $r=r_{\rm o}$, to 
a finite distance $r=a$, obeys the condition that the metric 
is continuous. Due to the spherical symmetry the components 
$g_{\theta\theta}$ and $g_{\phi\phi}$ are already continuous, 
and so one is left with imposing the continuity of 
$g_{tt}$ and $g_{rr}$, 
\begin{eqnarray} 
g_{tt({\rm int})}&=&g_{tt({\rm ext})} \,, \label{matchingmetrictt}\\ 
g_{rr({\rm int})}&=&g_{rr({\rm ext})} \,, \label{matchingmetricrr} 
\end{eqnarray} 
at $r=a$, with $g_{tt({\rm int})}$ and $g_{rr({\rm int})}$ being the 
metric components for the interior region at $r=a$, and $g_{tt({\rm 
ext})}$ and $g_{rr({\rm ext})}$ the exterior metric components 
for the vacuum solution at $r=a$. For the sake of consistency in 
the notation this could have been done in the orthonormal 
frame with the hat quantities $g_{\hat\mu\hat\nu}$, but in 
this case it is more direct to do with coordinate frame 
quantities. 
One can start the analysis by considering two general solutions 
of Equation 
(\ref{metricwormhole}), an interior solution and an exterior 
solution matched at a surface, $S$. The continuity of the metric 
then gives generically $\Phi_{\rm int}(a)=\Phi_{\rm 
ext}(a)$ and $b_{\rm int}(a)=b_{\rm ext}(a)$. If one now 
uses, Equations (\ref{metricwormhole}), 
(\ref{metricvacuumlambda}), (\ref{matchingmetrictt}) and 
(\ref{matchingmetricrr}),  one finds then 
$e^{2\Phi(a)}=\left [1-2GM/(c^2a)-\Lambda_{\rm ext}\,a^2/3 
\right]$ and $\left[1-b(a)/a \right]=\left 
[1-2GM/(c^2a)-\Lambda_{\rm ext}\,a^2/3 \right]$ which can be 
simplified to 
\begin{eqnarray} 
\Phi(a)&=&\frac{1}{2} \ln \left 
(1-\frac{2GM}{c^2a}-\frac{\Lambda_{\rm ext}}{3}a^2 \right)\,, 
\label{Phivalueata}\\ 
b(a)&=&\frac{2GM}{c^2}+\frac{\Lambda_{\rm ext}}{3}a^3 \,. 
\label{bvalueata} 
\end{eqnarray} 
From Equation (\ref{bvalueata}), one deduces that the mass of the 
wormhole is given by 
\begin{equation} 
M=\frac{c^2}{2G}\left(b(a)-\frac{\Lambda_{\rm ext}}{3}a^3 
\right)\,. \label{mass} 
\end{equation}

\bigskip 
{\it (ii) Matching of the equations I: The surface pressure} 
\medskip

We are going to consider the case where static interior 
observers measure 
zero tidal forces, i.e., $\Phi_{\rm int}=\,$constant,  
$\Phi_{\rm int}'=0$. 
This is, of course, a particular choice which simplifies the analysis.
As we have seen, the metric is continuous through the surface $S$.
However, their first and second derivatives might not be.  Since the
metric is static and spherically symmetric the only derivatives that
one needs to worry about are radial. Now, second derivatives in the metric 
are related to the Einstein tensor $G_{\mu\nu}$, or since 
we are working with hat quantities, to  $G_{\hat\mu\hat\nu}$. 
But $G_{\hat\mu\hat\nu}$ is proportional to the 
stress-energy tensor $T_{\hat\mu\hat\nu}$. Thus, something in the 
stress-energy tensor has to reflect this discontinuity. 
Indeed,  at the boundary $S$,  
$T_{\hat\mu\hat\nu}$ is proportional to a Dirac $\delta-$function, 
and we can write $T_{\hat\mu\hat\nu}=t_{\hat\mu\hat\nu}\, 
\delta(\hat{r}-\hat{a})$, where $\hat{r}=\sqrt{g_{rr}}\,r$ 
means the proper distance through the thin shell. 
To find $t_{\hat\mu\hat\nu}$ one then uses 
$\int_-^+G_{\hat\mu\hat\nu}\,d\hat{r}=
(8\,\pi\,G/c^4)
\int_-^+
t_{\hat\mu\hat\nu}\, 
\delta(\hat{r}-\hat{a})\,d\hat{r}$, where 
$\int_-^+$ means an infinitesimal integral through 
the shell. Using the property of the $\delta-$function 
$\delta\left(f(x)\right)=\left(1/|f'(x)|\right)\delta(x)$, 
and  $\int_-^+ g(x)\delta(x-x_0)=g(x_0)$, one finds 
\begin{equation}
t_{\hat\mu\hat\nu}= \frac{c^4}{8\,\pi\,G}
\int_-^+G_{\hat\mu\hat\nu}\,d\hat{r}\,.
\label{surfacestress}
\end{equation}
Since the shell is infinitesimally thin in the radial direction 
there is no radial pressure, thus we are left with a surface 
energy term $\Sigma$, and a surface tangential 
pressure ${\cal P}$.

First we calculate the surface energy density 
$\Sigma$. From Equation (\ref{Einsteintt}) 
we see that $G_{\hat t\hat t}$  
only depends on first derivatives of the metric, 
so that when integrated through the shell it will give 
metric functions only, that by definition are continuous. Thus, 
since the integral gives the value of the metric on the exterior 
side ($b^+$, say) minus the value of the metric on the interior 
side ($b^-$), it gives zero, and one finds
\begin{equation}
\Sigma=0\,.
\label{surfacenergy}
\end{equation}

Now we find the surface tangential pressure $\cal P$. 
From Equation (\ref{Einsteintheta}) 
we see that $G_{\hat\theta\hat\theta}$ has an 
important term $(1-\frac{b}{r})\Phi''$. The other 
terms depend at most on the first derivative and as before do 
no contribute to the integral. 
Thus, in this case Equation (\ref{surfacestress}) gives 
$8\pi\,G/c^2 {\cal P} = \sqrt{1-b(a)/a}\,\Phi'{_-^+}$.
Now, $\Phi'_-=0$ by  assumption, 
and $\Phi'^+= \left[G\,M/(c^2a^2)-\Lambda_{\rm ext}\,a/3\right]/
\left(1-b(a)/a\right)$. Thus, 
${\cal P}=\frac{c^4}{8\pi\,G\,a}
\frac{G\,M/(c^2a^2)-\Lambda_{\rm ext}\,a/3}{\sqrt{1-b(a)/a}}$, 
or more explicitly, 
\begin{equation} 
{\cal P} =\frac{c^4}{8\pi 
Ga}\,\frac{\frac{GM}{c^2a}-\frac{\Lambda_{\rm 
ext}}{3}a^2}{\sqrt{1-\frac{2GM}{c^2a}-\frac{\Lambda_{\rm 
ext}}{3}a^2}} \label{surfacetangentialpressure}\,. 
\end{equation} 
${\cal P}$ is always positive for the Schwarzschild and the 
Schwarzschild-anti de Sitter spacetime, i.e., $\Lambda_{\rm 
ext}\leq0$. The 
Schwarzschild-de Sitter spacetime, $\Lambda_{\rm 
ext}>0$, has to be analyzed more 
carefully. In Figure 3, we plot in a 
graph $9\Lambda_{\rm ext}(GM/c^2)^2\;\times \; 2GM/(c^2\,a)$ the 
regions where ${\cal P}$ is negative, zero or positive. For high 
$\Lambda_{\rm ext}(GMc^{-2})^2$ (either $\Lambda_{\rm ext}$ big or 
$M$ big) and at low $2GM/(c^2a)$ (either $M$ small or $a$ big), 
one needs a surface tension to support the structure. In the other 
case one needs a surface pressure. This is expected in the sense 
that for a positive $\Lambda_{\rm ext}$ one has an expanding 
external de Sitter spacetime. If $a$ is big (and so $2GM/(c^2a)$ small), the 
wormhole boundary is participating somehow in the expansion, so 
one needs a tension to hold it. For small $a$ ($\,\frac{2GM}{c^2a}\,$ big), 
the gravity wins over the expansion and so one needs a pressure to 
hold against collapse, a particular case being the  Schwarzschild case 
$\Lambda_{\rm ext}=0$. 
\vskip 0.5cm
\centerline{\epsffile{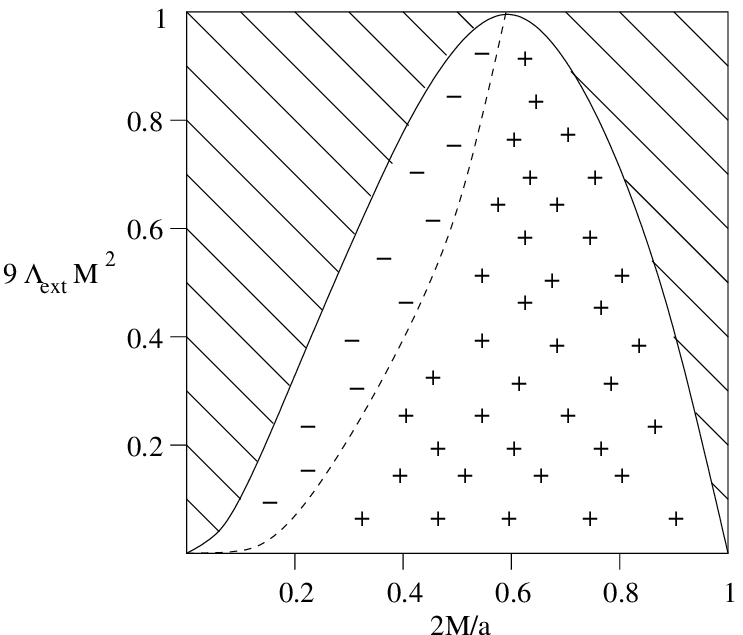}} 
\vskip 1mm
{\noindent {\small Figure 3 - 
The regions where $\cal P$ is negative and 
positive in a plot of $9\Lambda_{\rm ext}\,(GM/c^2)^2$
as a function of the inverse of the 
relative size of the wormhole, i.e., $2GM/c^2\,a$ 
(in the figure we have used geometrical units $G=1=c$) 
are given. 
Inside the 
solid line is the region of solutions. To the left of the 
dashed line ${\cal P}$ is a negative pressure, i.e., a tension, to 
the right ${\cal P}$ is a positive quantity, i.e., is a pressure,  
see text for more details.}} \label{fig3} 
\vskip 0.3cm
Now we detail the procedure to plot the graph. Define 
$\alpha =2GM/(c^2a)$ and $\beta =9\Lambda_{\rm ext}(GMc^{-2})^2$,  
so that $f=1-\frac{2GM}{c^2a}-\frac{\Lambda_{\rm ext}}{3}a^2$ 
can be written as $1-\alpha -\frac{4\beta}{27\alpha^{2}}$. The 
zero of $f$, defining the corresponding horizon for the 
set of parameters,  gives
$\beta$ as a function of $\alpha$. Varying 
$\alpha$ from $0$ to $1$ one gets the solid curve in Figure 3. 
The inside of this curve defines the region where ${\cal P}$ takes 
values (outside the curve there are no wormholes, 
on the top left hand part one mouth of the possible wormhole is causally 
disconnected from the other forever, and on the top right hand 
part the wormhole has been converted into a black hole ). 
We can check for the sign. To do this define 
$g=\frac{\alpha}{2}-\frac{4\beta}{27\alpha^{2}}$, see 
Equation (\ref{surfacetangentialpressure}). $g=0$ defines the curve 
$\beta=\frac{27}{8}\alpha^{3}$, which is plotted in the graph as a 
dashed line. To the left of this dashed curve, ${\cal P}$ is 
negative, and to the right it is positive.

One can have a term of comparison for the surface tangential 
pressure $\cal P$ at the thin shell. Assuming that the thin shell 
has a width of approximately $\Delta r$, one can 
consider a volumetric tangential pressure, orthogonal to the radial 
coordinate, given by 
\begin{equation}
\tilde{P}=\frac{{\cal P}}{\Delta r}\,.
\label{Pvolumetic}
\end{equation}
Taking into 
account Equation (\ref{pressureWHlambdaint}), we see that 
the 
tangential pressure at the mouth, with 
$\Phi'_{\rm int}=0$, is given by 
$\bar{p}(a)=\left(c^4/16\pi\,G\,a^3\right)
\left(b(a)-b'(a)a\right)$. 
Estimates of $\tilde{P}$ may be given in terms of 
$p(a)$, by defining the following ratio 
\begin{equation} 
\frac{\tilde{P}}{\bar{p}(a)}=
\,2\,\frac{a}{\Delta r}\,\frac{1}{\frac{b}{a}-b'}\,
\frac{ \frac{G\,M}{c^2a}-
\frac{\Lambda_{\rm ext}}{3}a^2}{\sqrt{1-\frac{2\,G\,M}{c^2}-
\frac{\Lambda_{\rm ext}}{3}\,a^3}}
\label{ptildeoverp0}\,. 
\end{equation} 
It is also interesting to find the ratio to $p(r_{\rm o})$, the maximum 
pressure, given by 
\begin{equation} 
\frac{\tilde{P}}{p(r_{\rm o})}=2\, \frac{r_{\rm o}^2}{a\,\Delta\,r}\,
\frac{1}{1-b'(r_{\rm o})}
\;\frac{\frac{GM}{c^2a}-\frac{\Lambda_{\rm 
ext}}{3}a^2}{\sqrt{1-\frac{2GM}{c^2a}-\frac{\Lambda_{\rm 
ext}}{3}a^2}}  \label{ptildeoverp}\,. 
\end{equation} 
One may find numerical estimates, considering various choices of 
the shape function, $b(r)$, which will be done while considering 
specific solutions of traversable wormholes.

\bigskip 
{\it (iii) Matching of the equations II: The radial pressure} 
\medskip

To construct specific solutions of wormholes with generic 
cosmological constant, one needs to know how the radial tension 
behaves across the junction boundary, $S$. The analysis is 
simplified if we consider two general solutions of Equation 
(\ref{metricwormhole}), an interior solution and an exterior 
solution matched at a surface, $S$. The radial component of 
the Einstein equations, Equation (\ref{tauWHlambda}), provides 
\begin{eqnarray} 
\frac{b_{\rm int}}{r^3}&=&\frac{8\pi G}{c^4}\tau_{\rm int} 
(r)+\Lambda_{\rm int}+2 \left(1-\frac{b_{\rm int}}{r} \right) 
\frac{\Phi'_{\rm int}}{r}  \label{gentauint} \,,\\ 
\frac{b_{\rm ext}}{r^3}&=&\frac{8\pi G}{c^4}\tau_{\rm ext} 
(r)+\Lambda_{\rm ext}+2 \left(1-\frac{b_{\rm ext}}{r} \right) 
\frac{\Phi'_{\rm ext}}{r}  \label{gentauext} \,. 
\end{eqnarray} 
Taking into account the continuity of the metric 
at the junction boundary one has obtained $\Phi_{\rm int}(a)=\Phi_{\rm 
ext}(a)$ and $b_{\rm int}(a)=b_{\rm ext}(a)$. For simplicity, we 
are considering $\Phi'_{\rm int}(a)=0$. Using again the 
following relation, 
$\Phi'_{\rm ext}(r)
=\left( \frac{GM}{c^2r^2}-\frac{\Lambda_{\rm ext}}{3}r \right)/
\left( 1-\frac{2GM}{c^2r}-\frac{\Lambda_{\rm ext}}{3}r^2\right)$, 
and taking into account  Equation (\ref{surfacetangentialpressure}), 
we verify that Equations (\ref{gentauint})-(\ref{gentauext}),
provide us with an 
equation which governs the behavior of the radial tension at the 
boundary, namely, 
\begin{equation} 
\tau_{\rm int} (a)+\frac{c^4}{8\pi G}\Lambda_{\rm int}=\tau_{\rm 
ext} (a)+\frac{c^4}{8\pi G} \Lambda_{\rm ext}+\frac{2}{a}{\cal P} 
e^{\Phi(a)} \label{radialtensionata}\, , 
\end{equation} 
where we have put $e^{\Phi(a)}=\sqrt{1-2GM/(c^2a)-\Lambda_{\rm 
ext} a^2/3}$. Equation (\ref{radialtensionata}), although 
not new in its most generic form \cite{Visser}, is a beautiful 
equation that relates the radial tension at the surface 
with the tangential pressure of the thin shell. 
A particularly interesting case is when ${\cal P} =0$.
In this situation $M=\Lambda_{\rm ext}c^2a^3/(3G)$, and 
$\Phi '_{\rm ext}(a)=0$. Since by our construction 
$\Phi '_{\rm int}(a)=0$, 
$\Phi '$ is continuous across the surface and the solution is 
reduced to a boundary surface. From Equation 
(\ref{bvalueata}) one finds that the shape 
function at the junction is given by 
$b(a)=\Lambda_{\rm ext}a^3$. Thus, Equation 
(\ref{radialtensionata}) simplifies to, 
$\tau_{\rm int} (a)+\frac{c^4}{8\pi G}\Lambda_{\rm int}=\tau_{\rm 
ext} (a)+\frac{c^4}{8\pi G}\Lambda_{\rm ext}$.  
If one considers a matching of an interior solution of a wormhole 
with generic $\Lambda$, given by Equation (\ref{metricwormhole}), to 
an exterior Schwarzschild solution, with $\tau_{\rm ext}=0$ and 
$\Lambda_{\rm ext}=0$, we simply have the condition that 
$\tau_{\rm int} (a)+\frac{c^4}{8\pi G}\Lambda_{\rm int}=0$ at the 
boundary surface. Matching an interior solution to an exterior 
Schwarzschild-de Sitter or Schwarzschild-anti de Sitter solution, 
with $\tau_{\rm ext}=0$ and $\Lambda_{\rm ext}\neq 0$, we have the 
following relationship, $\tau_{\rm int} (a)+\frac{c^4}{8\pi 
G}\Lambda_{\rm int}=\frac{c^4}{8\pi G}\Lambda_{\rm ext}$ at the 
boundary surface. These solutions will be analyzed in greater 
detail in the next sections.

\subsection{Spacetime diagrams} 
 
We now draw the spacetime diagrams, i.e., the Carter-Penrose 
diagrams, corresponding to wormholes in spacetimes with 
$\Lambda_{\rm ext}=0$, $\Lambda_{\rm ext}>0$, and 
$\Lambda_{\rm ext}<0$. 
They are easy to sketch once one knows the corresponding 
diagrams for the solution with no wormhole, i.e, 
the Carter-Penrose diagrams for Minkowski spacetime, 
de Sitter spacetime and anti-de Sitter spacetime 
\cite{hawkingellis}, respectively. Each point in the 
diagram represents a sphere. 

\vskip 0.3cm 
\centerline{\epsffile{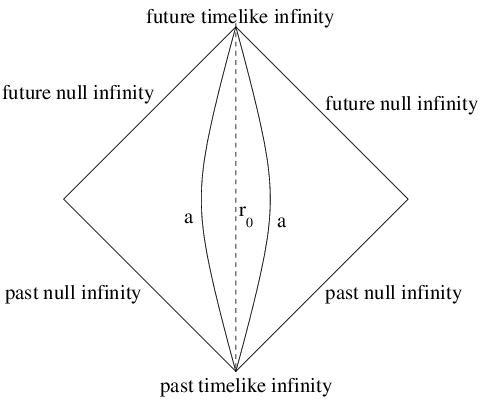}} 
\vskip 1mm 
{\noindent {\small Figure 4 - The spacetime diagram for the 
wormhole with $\Lambda_{\rm ext}=0$, i.e, a wormhole in an asymptotically 
Minkowski spacetime, represented by two copies of the Minkowski 
diagram joined at the throat.} } 
\vskip 0.3cm 

\centerline{\epsffile{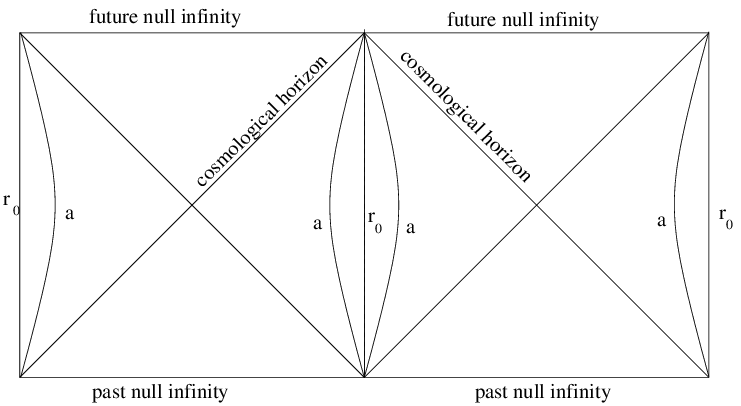}} 
\vskip 1mm 
{\noindent {\small Figure 5 - The spacetime diagram for the 
wormhole with $\Lambda_{\rm ext}>0$, 
i.e, a wormhole in an asymptotically de 
Sitter spacetime, with an infinite number of copies (only two are 
represented).}} 
\vskip 0.3cm 

\centerline{\epsffile{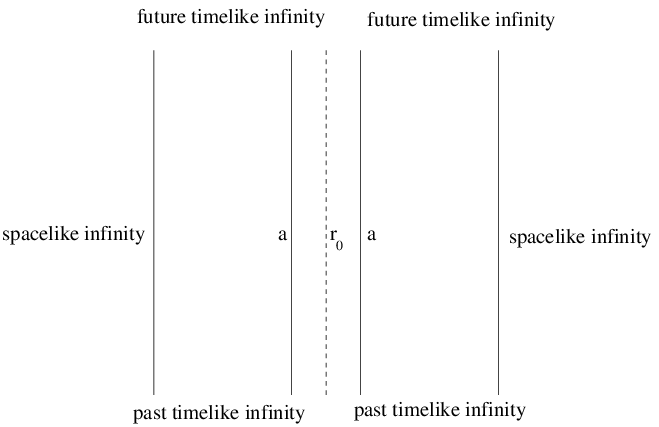}} 
\vskip 1mm 
{\noindent {\small Figure 6 - The spacetime diagram for 
the wormhole with $\Lambda_{\rm ext}<0$, i.e, a wormhole in an 
asymptotically anti-de Sitter spacetime, represented by two 
copies of the anti-de Sitter diagram joined at the throat.}} 
\vskip 0.3cm 

\noindent Since the wormhole creates an extra asymptotic region 
one has to duplicate the original diagram through the throat. 
In Figures 4, 5 and 6 the diagrams for a wormhole in an 
asymptotically flat spacetime $\Lambda_{\rm ext}=0$, 
in an asymptotically de Sitter spacetime $\Lambda_{\rm ext}>0$, 
and in an asymptotically anti-de Sitter spacetime 
$\Lambda_{\rm ext}<0$, respectively, are drawn. Note the 
duplication of the asymptotic regions.

\section{Specific construction of wormholes with \break generic 
$\Lambda$}

We will give some examples of traversable wormholes 
similar to those constructed in \cite{Morris}. The difference 
from the wormholes in that work is that the 
wormholes here in general have an infinitesimal 
thin shell with a tangential pressure 
${\cal P}\neq0$, and the exterior spacetime has a cosmological constant. 
We discuss first the case $\Lambda_{\rm ext}=0$, then 
$\Lambda_{\rm ext}>0$, and finally we mention briefly the 
case $\Lambda_{\rm ext}<0$. In all cases we put $\Phi'=0$ in the 
interior region.

\subsection{Specific solutions of traversable wormholes with 
$\Lambda_{\rm ext}=0$ (asymptotically flat wormholes)} 
 
\subsubsection{Matching to an exterior Schwarzschild solution, 
with ${\cal P}=0$}

Here we consider a matching of an interior solution with an 
exterior Schwarzschild solution ($\tau_{\rm ext}=0$ and 
$\Lambda_{\rm ext}=0$), and with the junction having 
zero tangential pressure, ${\cal P}=0$. From 
Equation (\ref{radialtensionata}) one has at the junction
\begin{equation} 
\tau_{\rm int} (a)+\frac{c^4}{8\pi G}\Lambda_{\rm int}=0 
\label{special}\,.
\end{equation} 
Then, from Equation (\ref{tauWHlambdaint}) (with $\Phi'=0$) 
one gets 
\begin{equation} 
0=\frac{c^4}{8\pi G}\frac{b(a)}{a^3} \,. 
\label{special2}\end{equation} 
Since $b\neq0$, Equation (\ref{special2})
is only satisfied if $a\rightarrow 
\infty$. This is one of the cases considered by Morris and Thorne 
\cite{Morris}, in which the wormhole's material extends from the 
throat all the way to infinity. 
 
\subsubsection{Matching to an exterior Schwarzschild solution, 
with ${\cal P} \neq 0$} 
 
Matching the interior solution to an exterior Schwarzschild 
solution ($\tau_{\rm ext}=0$ and $\Lambda_{\rm ext}=0$) but 
considering ${\cal P} \neq 0$, provides some interesting results. 
The behavior of the radial tension at the junction  is 
given by Equation (\ref{radialtensionata}) and  
taking into account Equation 
(\ref{bvalueata}) one finds that the shape function 
at the junction simply reduces to $b(a)=2GM/c^2$. We will next 
consider various choices for the shape function, $b(r)$, which will 
give different wormhole solutions.

\medskip 
{\it (i)} $b(r)=(r_{\rm o}\, r)^{1/2}$ 
\medskip
 
Consider the following functions 
\begin{eqnarray} 
\Phi(r)&=& \Phi _0 \,,\\ 
b(r)&=& (r_{\rm o} r)^{1/2} \,, 
\end{eqnarray} 
where $r_{\rm o}$ is the throat radius as before. Using the Einstein 
equations, Equations 
(\ref{rhoWHlambdaint})-(\ref{pressureWHlambdaint}) we have 
\begin{eqnarray} 
\bar\rho(r)\equiv\rho(r)+\frac{c^2}{8\pi G}\Lambda_{\rm int}&=& 
\frac{c^2}{8\pi G} \frac{r_{\rm o} ^{1/2}}{2r^{5/2}}
,\\ 
\bar\tau(r)\equiv
\tau(r)+\frac{c^2}{8\pi G}\Lambda_{\rm int}&=& \frac{c^4}{8\pi G} 
\frac{r_{\rm o} ^{1/2}}{r^{5/2}}\,,\\ 
\bar{p}(r)\equiv
p(r)+\frac{c^2}{8\pi G}\Lambda_{\rm int}&=& \frac{c^4}{8\pi G} 
\frac{r_{\rm o} 
^{1/2}}{4r^{5/2}}\, . 
\end{eqnarray} 
\noindent The distribution of the material threading the wormhole is
plotted in Figure 7.  
To find an estimate of the surface pressure at
the thin shell, one has $\frac{\tilde{P}}{p(r_{\rm o})}=\frac{4r_{\rm
o}^2GM}{\Delta r c^2a^2}\;\left(1-\frac{2GM}{c^2a}\right)^{-1/2}$ (see
Equation (\ref{ptildeoverp})).

From $b(a)=2GM/c^2$ and $b(a)=(r_{\rm o} a)^{1/2}$, one finds that 
the matching occurs at 
\begin{equation} 
a=\frac{\left(2GM/c^2\right)^2}{r_{\rm o}} \label{schwvaluea}\,. 
\end{equation} 
Now in order that the wormhole is not a black hole one has to 
impose $a>2GM/c^2$.  Then, from Equation (\ref{schwvaluea}) 
one finds $r_{\rm o}<2GM/c^2$. From Equation (\ref{schwvaluea}), we 
also extract the mass of the wormhole, given by 
$M=c^2\,(r_{\rm o} a)^{1/2}/(2G)$.

\centerline{\epsffile{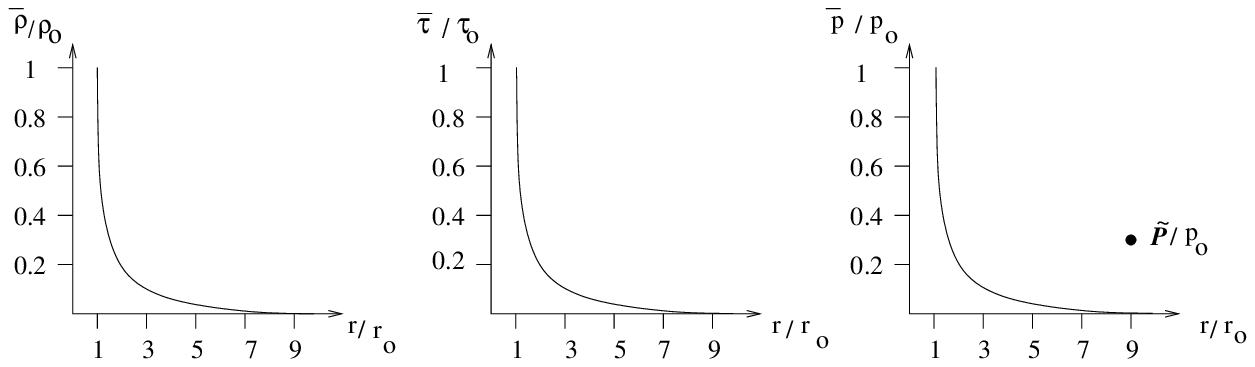}} 
{\noindent {\small Figure 7 - 
Distribution of the material threading the wormhole 
for the case $b(r)=(r_{\rm o}\, r)^{1/2}$. The 
mouth of the wormhole is at $a=9\,r_{\rm o}$. 
The averaged surface pressure $\tilde P$ on the thin shell 
defined in the text is depicted 
qualitatively as $\tilde P/p(r_{\rm o})$ in the 
${\bar p}/p(r_{\rm o})$ graph as a point.}} 
\vskip 0.5cm
The interior metric, $r_{\rm o}\leq r \leq a$, is determined recalling  
that $e^{2 \Phi(a)}=\left[1-\sqrt{r_{\rm o}/a} \right]$. 
It is given by 
\begin{equation} 
ds^2=-\left(1-\sqrt{\frac{r_{\rm o}}{a}} \right) 
\,c^2\,dt^2+\frac{dr^2}{\left(1-\sqrt{\frac{r_{\rm o}}{r}} 
\right)}+r^2\,(d\theta ^2+\sin^2{\theta}\, d\phi ^2) \,. 
\label{intscw} 
\end{equation} 
The exterior metric, $a\leq r < \infty$, is given by 
\begin{equation} 
ds^2=-\left(1-\frac{\sqrt{r_{\rm o} a}}{r} \right) 
\,c^2\,dt^2+\frac{dr^2}{\left(1-\frac{\sqrt{r_{\rm o} a}}{r} 
\right)}+r^2\,(d\theta ^2+\sin^2{\theta}\, d\phi ^2) \,. 
\label{extscw} 
\end{equation} 
The final metric of the whole spacetime is given by Equations 
(\ref{intscw})-(\ref{extscw}), which are joined smoothly, as we 
have carefully worked out. 
 
It is also interesting to briefly consider 
the traversability conditions that the absurdly advanced 
civilization might require 
to cross the wormhole from one mouth to the other and back (see 
the Appendix B for details). One finds that for an 
observer traversing the wormhole with a velocity  $v=0.01\,c$, the 
wormhole has a throat radius given by 
$r_{\rm o} \geq 500 \,{\rm km}$. One can  
choose $r_{\rm o}=500 \,{\rm km}$, and  assume 
that the traversal time done by the spaceship is approximately one 
year. Then, one finds that the matter distribution 
extends from $r_{\rm o}$ to $a=4.74\times 10^{13}\,{\rm m}\approx 5\times 
10^{-3}\,{\rm light \;years}$, with $a$ being the size of the wormhole. 
It is supposed that the space stations are parked there. 
One also finds that the wormhole 
mass is $M=3.3\times 10^{36}\,{\rm kg}$, six orders of 
magnitude superior to the Sun's mass. One may also find an 
estimate for Equation (\ref{ptildeoverp}), giving 
$\tilde{P}/p(r_{\rm o})\approx 10^{-4}$. 

One may choose other parameters, for instance,  
so that the wormhole mass is of the same 
order of the Sun's mass. Considering a traversal with a velocity 
$v=5.4\times 10^{3}\,{\rm m/s}$, we may choose that the 
wormhole throat is given by $r_{\rm o}=9\times 10^2 \,{\rm m}$. If we 
consider an extremely fast trip, where the traversal time is 
given by $\Delta \tau_{\rm traveler} =3.7\,{\rm s}$, the matter distribution 
extends from $r_{\rm o}$ to $a=10^{4}\,{\rm m}$. In this case the mass 
of the wormhole is given by $M\approx 2\times 10^{30}\,{\rm kg}$, 
which is the Sun's mass. From  Equation (\ref{ptildeoverp}) an 
estimate to $\tilde{P}$ 
is $\tilde{P}/p(r_{\rm o})\approx 5.7\times 10^3$. 
We have an extremely 
large surface pressure. As the wormhole mass is decreased, one sees 
that a larger tangential surface pressure is needed 
to support the structure.

\medskip 
{\it (ii)} $b(r)=r_{\rm o}^2/r$ 
\medskip 
 
Consider now, 
\begin{eqnarray} 
\Phi(r)&=& \Phi _0 \,,\\ 
b(r)&=& r_{\rm o}^2/r \,.
\end{eqnarray} 
Using the Einstein equations, Equations 
(\ref{rhoWHlambdaint})-(\ref{pressureWHlambdaint}), one has 
\begin{eqnarray} 
\bar\rho(r)\equiv\rho(r)+
\frac{c^2}{8\pi G}\Lambda_{\rm int}&=& -\frac{c^2}{8\pi G} 
\frac{r_{\rm o} ^2}{r^4}
   \,,\\ 
\bar\tau(r)\equiv\tau(r)+
\frac{c^2}{8\pi G}\Lambda_{\rm int}&=& \frac{c^4}{8\pi G} 
\frac{r_{\rm o} ^2}{r^4}
   \,,\\ 
\bar p(r)\equiv p(r)+\frac{c^2}{8\pi G}\Lambda_{\rm int}&=& 
\frac{c^4}{8\pi G} 
\frac{r_{\rm o} ^2}{r^4}\,. 
\end{eqnarray} 
The distribution of the material threading the wormhole is plotted 
in Figure 8. A qualitative estimate for ${\cal P}/p(r_{\rm o})$, 
in this case given by $\frac{\tilde{P}}{p(r_0)}=\frac{4r_0^2GM}{\Delta r
c^2a^2}\;\left(1-\frac{2GM}{c^2a}\right)^{-1/2}$ (see Equation
(\ref{ptildeoverp})) is 
also plotted. 

From $b(a)=2GM/c^2$ and $b(r)=r_{\rm o}^2/r$, the matching occurs at 
\begin{equation} 
a=\frac{c^2r_{\rm o}^2}{2GM}\,. 
\end{equation} 
Imposing $a>2GM/c^2$, so that the wormhole does not correspond to 
a black hole solution, we have $r_{\rm o}>2GM/c^2$ and the mass is given 
by $M=\frac{c^2r_{\rm o}^2}{2Ga}$.

\noindent The interior metric, $r_{\rm o}\leq r \leq a$, with 
$e^{2\Phi(a)}=\left[1-(r_{\rm o}/a)^2\right]$, is given by 
\begin{equation} 
ds^2=-\left[1-\left (\frac{r_{\rm o}}{a}\right)^2 \right] 
\,c^2\,dt^2+\frac{dr^2}{1-\frac{r_{\rm o}^2}{r^2}}+r^2\,(d\theta 
^2+\sin^2{\theta}\, d\phi ^2) \,. \label{secondintscw} 
\end{equation} 
The exterior metric, $a\leq r < \infty$, is given by 
\begin{equation} 
ds^2=-\left(1-\frac{r_{\rm o}^2}{ar} \right) 
\,c^2\,dt^2+\frac{dr^2}{\left(1-\frac{r_{\rm o}^2}{ar} 
\right)}+r^2\,(d\theta ^2+\sin^2{\theta}\, d\phi ^2) \,. 
\label{secondextscw} 
\end{equation} 
The respective final metric solution of the spacetime is given by 
Equations (\ref{secondintscw})-(\ref{secondextscw}), joined 
smoothly at $a$. 
 
\vskip 0.5cm 
\centerline{\epsffile{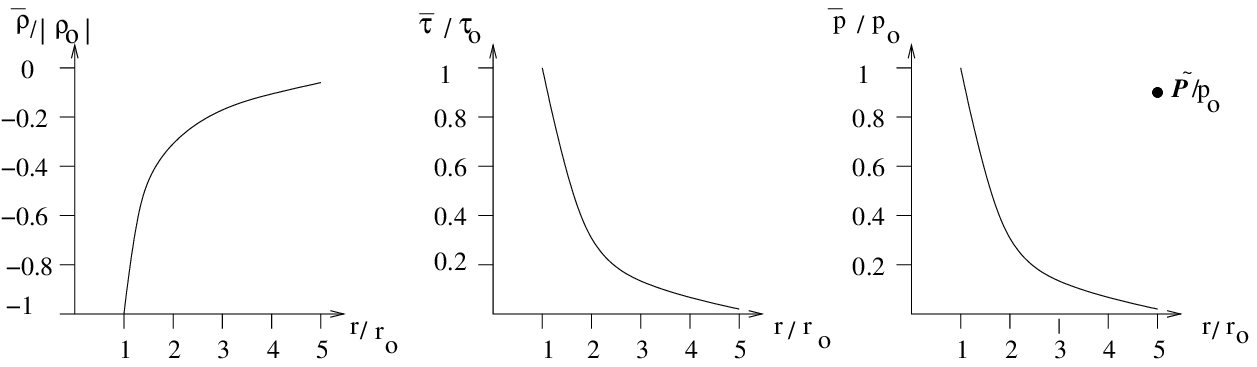}} 
{\noindent {\small Figure 8 - 
Distribution of the material threading the wormhole for 
the case  $b(r)=r_{\rm o}^2/r$.  The 
mouth of the wormhole is at $a=5\,r_{\rm o}$.
$\tilde{P}/p(r_{\rm o})$ is depicted qualitatively in the 
${\bar p}/p(r_{\rm o})$ graph as a point.}} 
\vskip 0.5cm

A comment on the interior metric (\ref{secondintscw}) is in order. 
In fact, this interior solution is the same as the one found by 
Ellis \cite{homerellis}. Indeed, considering the following coordinate 
transformations, $\bar{t}=\left[1-\left(r_{\rm o}/a \right)^2 
\right]^{1/2}t$ and $l=\pm \left(r^2-r_{\rm o}^2 \right)^{1/2}$, one 
finds that the metric is reduced to 
$ds^2=-c^2\,d\bar{t}\,^2+dl^2+\left(r_{\rm o}^2+l^2 \right)\,(d\theta 
^2+\sin^2{\theta}\, d\phi ^2) \,,$ where $l$ is the proper radial 
coordinate, ranging from $-\infty$ to $+\infty$. The properties 
are commented in \cite{Morris} and \cite{clement1}. Harris showed 
further that it is a solution of the Einstein equations 
with a stress-energy tensor of a 
peculiar massless scalar field \cite{harris}.

With respect to the traversability conditions given in the 
Appendix B, considering that the traversal velocity is 
$v=0.01\,c$, the wormhole throat is $r_{\rm o}=10^6\,{\rm m}$. If the 
traversal time is one year, $\Delta \tau_{\rm traveler}=3.16\times 
10^7 \,{\rm s}$, the junction is at $a=4.74\times 10^{13}\,{\rm 
m}\approx 5\times 10^{-3}\,{\rm light\; years}$. The mass of the 
wormhole is $M\approx 1.4 \times 10^{25} \,{\rm kg}$, 
approximately an order of magnitude superior to the Earth's mass. 
An estimate for Equation (\ref{ptildeoverp}), is given by 
$\tilde{P}/p(r_{\rm o})\approx 4.6\times 10^{-16}$.

\subsection{Specific solutions of traversable wormholes with 
$\Lambda_{\rm ext}>0$ (asymptotically de Sitter wormholes) }

\subsubsection{Matching to an exterior Schwarzschild-de Sitter 
solution, with ${\cal P} =0$} 
 
In this section we will be interested in a matching of an interior 
solution with an exterior Schwarzschild-de Sitter solution, 
$\tau_{\rm ext}=0$ and $\Lambda_{\rm ext}> 0$, at a boundary 
surface, ${\cal P} =0$. We verify from Equation 
(\ref{radialtensionata}) that the following condition holds 
\begin{equation} 
\tau_{\rm int} (a)+\frac{c^4}{8\pi G}\Lambda_{\rm 
int}=\frac{c^4}{8\pi G}\Lambda_{\rm ext}\,, 
\end{equation} 
at the surface boundary. Considering Equation 
(\ref{tauWHlambdaint}), we have 
\begin{equation} 
b(a)=\Lambda_{\rm ext}\,a^3 \label{bata}\,. 
\end{equation} 
Substituting this value in Equation (\ref{mass}), one obtains the 
mass of the wormhole, given by, 
\begin{equation} 
M=\frac{c^2}{3G}\,\Lambda_{\rm ext}\,a^3 \,. 
\end{equation} 
We shall next consider identical shape functions as in the above 
section. 
 
\medskip 
{\it (i)} $b(r)=(r_{\rm o}\, r)^{1/2}$ 
\medskip 
 
Consider the following functions 
\begin{eqnarray} 
\Phi(r)&=& \Phi _0 \,,\\ 
b(r)&=& (r_{\rm o} r)^{1/2} \,.
\end{eqnarray} 
From $b(a)=\Lambda_{\rm ext}\,a^3$ and $b(a)=(r_{\rm o} a)^{1/2}$, one sees 
that the matching occurs at 
\begin{equation} 
a\equiv \frac{r_{\rm o}^{1/5}}{\Lambda_{\rm ext} ^{2/5}}\,. 
\end{equation} 
The mass can then be expressed as $M=c^2(r_{\rm o}a)^{1/2}/(3G)$.
It can be shown that the interior solution, $r_{\rm o}\leq r \leq a$, is 
identical to Equation (\ref{intscw}), i.e., 
\begin{equation} 
ds^2=-\left(1-\sqrt{\frac{r_{\rm o}}{a}} \right) 
\,c^2\,dt^2+\frac{dr^2}{\left(1-\sqrt{\frac{r_{\rm o}}{r}}
\right)}+r^2\,(d\theta ^2+\sin^2{\theta}\, d\phi ^2) \,. 
\label{intdesitter} 
\end{equation} 
The exterior solution, $a\leq r < \infty$, is given by the 
following metric 
\begin{eqnarray} 
&ds^2=-\left(1-\frac{2(r_{\rm o}a)^{1/2}}{3r}-
\frac{r_{\rm o}^{1/2}r^2}{3a^{5/2}} 
\right) \,c^2\,dt^2+\frac{dr^2}{\left(1-\frac{2(r_{\rm o}a)^{1/2}}{3r}- 
\frac{r_{\rm o}^{1/2}r^2}{3a^{5/2}} \right)}\nonumber\\
&
+r^2\,(d\theta 
^2+\sin^2{\theta}\, d\phi ^2) \,. \label{extdesitter} 
\end{eqnarray} 
The spacetime of the final solution is given by the metrics, 
Equations (\ref{intdesitter})-(\ref{extdesitter}), which have been 
smoothly joined at $a$.

The additional parameter now is the cosmological constant, 
$\Lambda_{\rm ext}$, given by $\Lambda_{\rm ext}=(r_{\rm o}/a^5)^{1/2}$. 
For instance, consider a traversal velocity  $v=0.01\,c$, so 
that $r_{\rm o}=5\times 10^5 \,{\rm m}$. If the observer traverses 
through the wormhole comfortably during a year, $\Delta 
\tau_{\rm traveler}\approx 3.16\times 10^7 \,{\rm s}$, and 
$a=4.74\times 
10^{13}\,{\rm m}$. The mass of the wormhole 
is $M\approx 2.2 \times 10^{36} \,{\rm 
kg}$ and the cosmological constant has the value $\Lambda_{\rm 
ext}=4.6 \times 10^{-32}\,{\rm m^{-2}}$. The 
cosmological event horizon is then situated at $r_c=8.1\times 
10^{15}\,{\rm m}\approx\,200\,a$.

\medskip 
{\bf (ii) $b(r)=r_{\rm o}^2/ r$} 
\medskip 
 
Consider now the functions 
\begin{eqnarray} 
\Phi(r)&=& \Phi _0 \,,\\ 
b(r)&=& r_{\rm o}^2/ r \,. 
\end{eqnarray} 
From $b(a)=\Lambda_{\rm ext}\,a^3$ and $b(a)=r_{\rm o}^2/a$ one sees 
that the matching occurs at $a=r_{\rm o}^{1/2}/\Lambda_{\rm 
ext}^{1/4}$. The mass can be written as $M=c^2r_{\rm o}^2/(3Ga)$. 
The exterior cosmological constant is then 
$\Lambda_{\rm ext}=(r_{\rm o}/a^2)^2$. 

Considering that the traversal velocity is $v=0.01\,c$, the throat
radius is $r_{\rm o}=10^6\,{\rm m}$. If the traversal time is done in one
year the matching occurs at $a=4.74\times 10^{13}\,{\rm m}$. The mass
is then given by $M=9.5\times 10^{24}\,{\rm kg}$, the same order of
magnitude as the Earth's mass, and the cosmological constant is
$\Lambda_{\rm ext}=1.98\times 10^{-43}\,{\rm m^{-2}}$.  The
cosmological horizon is situated at $r_c=3.9\times 10^{21}\,{\rm
m}\approx\,10^8\,a$.

Another example is provided by a traveler 
with velocity is $v=0.0001\,c$. The throat radius is then 
$r_{\rm o}=10^4\,{\rm m}$. Assuming that the traversal time is done in 
$\Delta \tau_{\rm traveler}=6.3\times 10^3\,{\rm years}$, 
one has $a\approx 
3\times 10^{15}\,{\rm m} \approx 0.32$ light years. The mass is 
given by $M=1.5\times 10^{19}\,{\rm kg}$, the mass of an asteroid, 
and the 
cosmological constant has a value of $\Lambda_{\rm 
ext}=10^{-54}\,{\rm m^{-2}}$, approximately the present value.

\subsubsection{Matching to an exterior Schwarzschild-de Sitter 
solution, with ${\cal P} \neq 0$} 
 
One can also match the interior solution with an exterior 
Schwarzschild-de Sitter solution ($\tau_{\rm ext}=0$ and 
$\Lambda_{\rm ext}> 0$) in the presence of a thin shell, ${\cal P} 
\neq 0$. From Equation (\ref{radialtensionata}), we have the 
behavior of the radial tension at the thin shell, given by 
\begin{equation} 
\tau_{\rm int} (a)+\frac{c^4}{8\pi G}\Lambda_{\rm 
int}=\frac{c^4}{8\pi G}\Lambda_{\rm ext}+\frac{2}{a}{\cal P} 
\,e^{\Phi(a)}\,. 
\end{equation} 
The shape function at the junction is given by Equation 
(\ref{bvalueata}). From Equation (\ref{mass}), one verifies that 
the mass of the wormhole is zero when $b(a)=\Lambda_{\rm ext} 
a^3/3$, is positive when $b(a)>\Lambda_{\rm ext} a^3/3$, and 
is negative when $b(a)<\Lambda_{\rm ext} a^3/3$. One can perform 
a similar analysis as done for the previous examples.

\subsection{Specific solutions of traversable wormholes with 
$\Lambda_{\rm ext}<0$ (asymptotically anti-de Sitter wormholes)}

\subsubsection{Matching to an exterior Schwarzschild-anti de Sitter 
solution, with ${\cal P} =0$} 
 
From Equation 
(\ref{radialtensionata}), matching an interior solution 
with an exterior given by the Schwarzschild-anti 
de Sitter solution ($\tau_{\rm ext}=0$ and $\Lambda_{\rm ext}< 0$), 
at a boundary surface ${\cal P} =0$, yields, 
\begin{equation} 
\tau_{\rm int} (a)+\frac{c^4}{8\pi G}\Lambda_{\rm 
int}=-\frac{c^4}{8\pi G}\left|\Lambda_{\rm ext}\right|\,, 
\end{equation} 
at the surface boundary. Considering Equation 
(\ref{tauWHlambdaint}), we have 
\begin{equation} 
b(a)=-\left|\Lambda_{\rm ext}\right|\,a^3 \,. 
\end{equation} 
From Equation (\ref{lift}), we concluded that the shape function 
has to be positive to guarantee that the factor $\sqrt{r/b-1}$ be real. 
Therefore, for the anti-de Sitter exterior, i.e., 
$\Lambda_{\rm ext}<0$,  with ${\cal P}=0$ 
there is no solution. This problem  may be 
overcome by considering a matching to an exterior anti-de Sitter 
solution with a thin shell, i.e., ${\cal P} \neq 0$.

\subsubsection{Matching to an exterior Schwarzschild-anti de Sitter 
solution, with ${\cal P} \neq 0$} 
 
From Equation (\ref{bvalueata}), one finds that 
$b(a)$ is positive if 
\begin{equation} 
\frac{2GM}{c^2}\geq \frac{|\Lambda_{\rm ext}|}{3}a^3\,. 
\end{equation} 
Then one can construct easily wormholes in anti-de Sitter spacetime, 
and again perform a
similar analysis as done for the previous examples.

\section{Conclusions}

We have considered Morris$\,$-Thorne wormholes, i.e., static and
spherically symmetric traversable wormholes, in the presence of a
non-vanishing cosmological constant. Matching the interior solution
with a vacuum exterior solution, we have deduced an equation for the
tangential surface pressure, and another one which governs the
behavior of the radial tension at the boundary.
 
Specific solutions with various choices of the shape function were
presented.  Through the traversability conditions, we have obtained
estimates for the matching boundary, $a$, the mass of the wormhole $M$,
and the tangential surface pressure $\cal P$, by imposing values for
the traversal velocity and the traversal time.

\vskip 1.5cm
 
\noindent {\bf Acknowledgements}\hskip0.2cm 
JPSL thanks Joseph Katz and Donald Lynden-Bell
for teaching many years ago how to do junctions in an easy way, 
thanks Madalena Pizarro for alerting that a civilization that
constructs a wormhole is an absurdly advanced civilization,
rather than arbitrarily advanced, and
thanks Observat\'orio Nacional do Rio de Janeiro for hospitality. FSNL
thanks many conversations with Paulo Crawford do Nascimento.  The
present address of SQO is Instituto de F\'{\i}sica, Universidade
Federal do Rio de Janeiro, CEP 21945-970, Rio de Janeiro.
This work was partially funded through project PESO/PRO/2000/4014 by
Funda\c c\~ao para a Ci\^encia e Tecnologia (FCT) -- Portugal. 
 
\newpage 

\appendix 
 
\section{Solutions of 
$(1-\frac{2\,G\,M}{c^2\,r}-\frac{\Lambda_{\rm ext}}{3}\,r^2)=0$} 
 
The roots of the cubic equation
\begin{equation} 
x^3+ax+b=0  \label{cubic} \,, 
\end{equation} 
can be found. Indeed, defining 
\begin{eqnarray} 
A&=&\sqrt[3]{-\frac{b}{2}+\sqrt{\frac{b^2}{4}+\frac{a^3}{27}}}  \,, \\ 
B&=&\sqrt[3]{-\frac{b}{2}-\sqrt{\frac{b^2}{4}+\frac{a^3}{27}}} \,, 
\end{eqnarray} 
the solutions are given by 
\begin{eqnarray} 
x_1&=&A+B    \label{root1}      \,, \\ 
x_2&=&-\frac{A+B}{2}+\frac{A-B}{2}\sqrt{-3}  \label{root2}  \,, \\ 
x_3&=&-\frac{A+B}{2}-\frac{A-B}{2}\sqrt{-3}    \label{root3} 
\,. 
\end{eqnarray}

\bigskip 
\noindent 
{\it (i) The Schwarzschild spacetime, $\Lambda_{\rm 
ext}=0$:} 

\noindent 
This case is trivial and is analyzed directly in the 
text.

\bigskip 
\noindent 
{\it (ii) The Schwarzschild-de Sitter spacetime, $\Lambda_{\rm 
ext}>0$:} 
 
\noindent 
The equation $f(r)=(1-\frac{2GM}{c^2r}-\frac{\Lambda_{\rm 
ext}}{3}r^2)=0$ may be recast into the form 
\begin{equation} 
x^3-\sigma x+\sigma=0 \label{cubicSchwdS}\,, 
\end{equation} 
with $x=r/(2GMc^{-2})$ and $\sigma =3c^4/(4\Lambda_{\rm 
ext}G^2M^2)$. Thus comparing (\ref{cubicSchwdS}) with 
(\ref{cubic}) one finds 
\begin{eqnarray} 
A&=&\sqrt[3]{-\frac{\sigma}{2}+\sqrt{\frac{\sigma^2}{4}-
\frac{\sigma^3}{27}}} 
\label{A}   \,, \\ 
B&=&\sqrt[3]{-\frac{\sigma}{2}-\sqrt{\frac{\sigma^2}{4}-
\frac{\sigma^3}{27}}} 
\label{B}  \,. 
\end{eqnarray} 
To find which two of the three solutions one 
must pick up, we consider the case 
$\Lambda_{\rm ext}(GM/c^2)^2 \ll 1$, i.e., $\sigma \gg 1$. The square 
root $\sqrt{\frac{\sigma^2}{4}-\frac{\sigma^3}{27}}$ may be 
expanded as 
$i\,\frac{\sigma^{3/2}}{\sqrt{27}}\left(1-\frac{27}{8\sigma}\right)$. 
Thus Equations (\ref{A})-(\ref{B}) yield the following 
approximation 
\begin{eqnarray} 
A&=&\sqrt[3]{i}\;\frac{\sigma^{1/2}}{\sqrt{3}}\left(1+i\,
\frac{\sqrt{27}}{6}\frac{1}{\sigma 
^{1/2} }\right)   \label{approxA}  \,,  \\ 
B&=&\sqrt[3]{-i}\;\frac{\sigma^{1/2}}{\sqrt{3}}\left(1-i\,
\frac{\sqrt{27}}{6}\frac{1}{\sigma 
^{1/2} }\right)   \label{approxB}  \,. 
\end{eqnarray} 
Defining $\alpha =\frac{\sigma^{1/2}}{\sqrt{3}}$ and 
$\beta=\frac{\sqrt{27}}{6}\frac{1}{\sigma^{1/2}}$, and considering 
$\sqrt[3]{i}=e^{i\,\frac{\pi}{6}}=\frac{\sqrt{3}}{2}+\frac{i}{2} 
\,$,
$\sqrt[3]{-i}=e^{-i\,\frac{\pi}{6}}=\frac{\sqrt{3}}{2}-\frac{i}{2} 
\,$, Equations (\ref{approxA})-(\ref{approxB}) take the form 
\begin{eqnarray} 
A&=&\alpha \left[\left(\frac{\sqrt{3}}{2}-\frac{\beta}{2} 
\right)+i\,\left(\frac{1}{2}+\frac{\sqrt{3}}{2}\beta 
\right)\right]  \label{finalA} 
\,, \\ 
B&=&\alpha \left[\left(\frac{\sqrt{3}}{2}-\frac{\beta}{2} 
\right)-i\,\left(\frac{1}{2}+\frac{\sqrt{3}}{2}\beta 
\right)\right]   \label{finalB}  \,. 
\end{eqnarray} 
We are only interested in the positive solutions. Thus, 
substituting Equations (\ref{finalA})-(\ref{finalB}) into 
Equations (\ref{root1})-(\ref{root3}) and taking into account 
$x=r/(2GMc^{-2})$, we have that $x_3$ and $x_1$ give, respectively 
\begin{eqnarray} 
r_b&=&\frac{2GM}{c^2}\left[1+\frac43
\Lambda_{\rm ext}\left(\frac{G\,M}{c^2}\right)^2\right] \label{sol1} 
\,, \\
r_c&=&\sqrt{\frac{3}{\Lambda_{\rm ext}}} 
\left(1-\frac{GM}{c^2}\sqrt{\frac{\Lambda_{\rm ext}}{3}}\right) 
  \label{sol2} \,. 
\end{eqnarray} 
We see that $r_b$ and $r_c$ correspond to the vacuum black hole horizon 
and to the 
cosmological event horizon, respectively. In Equation 
(\ref{sol1}) to find the numerical factor of the first 
order term $\Lambda_{\rm ext}\left(\frac{G\,M}{c^2}\right)^2$ it 
is easier to linearize the solution by writing $x=1+\epsilon$, for 
small $\epsilon$,  
and then with the help of (\ref{cubicSchwdS}) one finds $\epsilon$.

\bigskip 
\noindent 
{\it (iii) The Schwarzschild-anti de Sitter spacetime, 
$\Lambda_{\rm ext}<0$:} 
 
\noindent In the Schwarzschild-anti de Sitter spacetime, 
one finds that the 
equation $f(r)=(1-\frac{2GM}{c^2r}+\frac{|\Lambda_{\rm 
ext}|}{3}r^2)=0$  may be recast into the form 
\begin{equation} 
x^3+\sigma x-\sigma=0 \label{cubicSchwAdS}\,, 
\end{equation} 
with $x=r/(2GMc^{-2})$ and $\sigma =3c^4/
(4|\Lambda_{\rm ext}|\,G^2M^2)$,
$|\Lambda_{\rm ext}|\equiv-\Lambda_{\rm ext}$. Thus comparing 
(\ref{cubicSchwAdS}) with 
(\ref{cubic}) one finds 
\begin{eqnarray} 
A&=&\sqrt[3]{\frac{\sigma}{2}+
\sqrt{\frac{\sigma^2}{4}+\frac{\sigma^3}{27}}} 
 \label{Aads}   \,, \\ 
B&=&\sqrt[3]{\frac{\sigma}{2}-
\sqrt{\frac{\sigma^2}{4}+\frac{\sigma^3}{27}}} 
\label{Bads}  \,. 
\end{eqnarray} 
The only real solution is $x_1=A+B$. Indeed, for 
$|\Lambda_{\rm ext}|\,(GM/c^2)^2 \ll 1$, i.e., $\sigma \gg 1$ one finds 
\begin{equation} 
r_b=\frac{2GM}{c^2}\left[1-\frac43
|\Lambda_{\rm ext}|\,\left(\frac{G\,M}{c^2}\right)^2\right] \,, 
\label{solAdS} 
\end{equation} 
which gives the black hole horizon. In Equation 
(\ref{solAdS}) to find the numerical factor of the first 
order term $\Lambda_{\rm ext}\left(\frac{G\,M}{c^2}\right)^2$ it 
is easier to linearize the solution by writing $x=1+\epsilon$, for 
small $\epsilon$,  
and then with the help of (\ref{cubicSchwAdS}) one finds $\epsilon$.

\section{Traversability conditions} 
 
We will be interested in specific solutions for traversable 
wormholes and assume that a traveler of an absurdly advanced 
civilization, with human traits, begins the trip in a space 
station in the lower universe, at proper distance 
$l=-l_1$, and ends up in the 
upper universe, at $l=l_2$. We shall, for self-containment and 
self-consistency, briefly describe the traversability conditions 
given in \cite{Morris}. 
 
The cosmological constant does not enter the analysis directly.
Indeed, although the interior cosmological constant, $\Lambda_{\rm
int}$, can be incorporated into the effective quantities, for
$\Lambda_{\rm ext}\neq 0$ the external parameters such as the mass of
the wormhole change.  Thus, the cosmological constant enters the
traversability conditions indirectly. Ford and Roman \cite{fordroman2}
with their quantum inequalities have imposed severe restrictions on the
$T_{\mu\nu}$ obeyed by matter, and in particular found that the
parameters of the Morris$\,$-Thorne$\,$-$\,$Yurtsever 
\cite{mty} wormhole obeyed
these inequalities. However, we now know that the energy conditions can
be even classically violated \cite{barcelovisser1}, and thus the
Ford-Roman inequalities have somehow lost their strength in this
context. Therefore, our parameters are not chosen in order to satisfy
the quantum inequalities, but rather follow the spirit of the Morris
and Thorne work \cite{Morris} where attention is paid to the
traversability conditions of a human being.
 
\bigskip 
{\it (i) Tidal acceleration felt by a traveler}
 
\medskip 
 
The tidal accelerations between two parts of the traveler's body, 
separated by, say, 2 meters,  
should not exceed the gravitational acceleration at 
Earth's surface $g_{\rm Earth}$, 
(with $g_{\rm Earth}\approx10$m/s$^2$).
From \cite{Morris} and using our simplifying assumption 
$\Phi'=0$, one obtains the following 
inequality
\begin{equation} 
\left| \frac{\gamma ^2}{2r^2} \left(\frac{v}{c}\right)^2 
\left (b'-\frac{b}{r}\right ) \right|\leq 
\frac{g_{\rm Earth}}{2c^2\, 1{\rm m}}\approx \frac{1}{10^{16}\,{\rm m}^2} 
\label{tidal2}\,,  
\end{equation} 
with $\gamma=(1-v^2/c^2)^{-1/2}$, and $v$ being the traveler's velocity. 
This inequality refers to tangential tidal accelerations. 
Radial tidal acceleration are zero for $\Phi'=0$.
From Equation (\ref{tidal2}) one sees that stationary 
observers with $v=0$ measure null tidal forces.

\bigskip 
{\it (ii) Acceleration felt by a traveler}
 
\medskip 
 
Another condition that needs to be respected is that the local 
gravitational acceleration, $|\vec{a}|$ at the space stations 
should not surpass the terrestrial gravitational acceleration, 
$g_{\rm Earth}$. The condition imposed in \cite{Morris} 
(for $\Phi'=0$) is 
\begin{equation} 
|\vec{a}|=\left |\left (1-\frac{b}{r}\right)^{1/2} 
\gamma'c^2 \right|\leq g_{\rm Earth} 
\label{acceleration} \,. 
\end{equation} 
For $v=\,$constant travelers, one has $|\vec{a}|=0$, of course!

\bigskip 
{\it (iii) Total time in a traversal}
 
\medskip 
 
The trip should take a relatively short time, for instance one 
year, as measured by the traveler and for observers that stay at 
rest at the space stations, $l=-l_1$ and $l=l_2$, i.e., 
\begin{eqnarray} 
\Delta \tau_{\rm traveler} &=&\int_{-l_1}^{+l_2} \frac{dl}{v\gamma} \leq 
1\,\,{\rm year} 
\label{travelertime}, \\ 
\Delta t_{\rm space\,station} 
&=&\int_{-l_1}^{+l_2} \frac{dl}{v e^{\Phi}} \leq 
1\,\,{\rm year} \label{observertime}, 
\end{eqnarray} 
respectively. 
 
Having set these conditions for general wormholes, 
we will now study two particular cases.

\bigskip \bigskip
\noindent {\bf (1) The wormhole with shape function 
$b(r)=(r_{\rm o} r)^{1/2}$} 
\medskip 

{\it (i) Tidal acceleration felt by a traveler}
 
In the interior region, $r_{\rm o}<r<a$,  the tidal acceleration as 
measured by a traveler moving radially through the wormhole is
given by Equation (\ref{tidal2}).  
For non-relativistic velocities, $v\ll c$, we have $\gamma \approx 
1$, and substituting the expression of $b(r)$, i.e., $b(r)=(r_{\rm o} 
r)^{1/2}$, in Equation (\ref{tidal2}), one can impose
a velocity for the traveler traversing through the wormhole, such 
that the tidal accelerations felt are inferior to the 
terrestrial gravitational acceleration, $g_{\rm Earth}$. 
With these conditions, one obtains a restriction for the 
velocity at the throat, where 
the acceleration is severest, given by 
\begin{equation} 
\left(\frac{v}{c}\right)\leq \frac{2r_{\rm o}}{10^8 {\rm m}} \,.
\end{equation} 
If the observer traverses the wormhole with a non-relativistic 
velocity $v\ll c$, the accelerations at the beginning and end of 
the trip are negligible. If we consider that the velocity is $v 
\approx 0.01 \,c$, one can find estimates for the dimensions of 
the wormhole. The wormhole throat obeys $r_{\rm o} \geq 500 \,{\rm km}$. 
For definiteness consider the choice $r_{\rm o} = 500 \,{\rm km}$. 
Taking into account Equation (\ref{schwvaluea}) for 
asymptotically flat spacetimes, the region of 
matter distribution will extend to 
\begin{equation} 
a=\frac{(2GM/c^2)^2}{(500{\rm km})}\,. 
\end{equation} 
Then, by choosing $a$, we find the value of the wormhole mass. We 
now take the steps to choose $a$. Similar procedures 
follow for $\Lambda_{\rm ext}\neq0$, see main text.

\bigskip 
{\it (ii) Acceleration felt by a traveler}
 
Another condition that needs to be respected is that the local 
gravitational acceleration $|\vec{a}|$, at the space stations 
should not surpass the terrestrial gravitational acceleration, 
$g_{\rm Earth}$. Considering non-relativistic velocities, $v\ll c$, 
and $v\approx\,$constant, one has $|\vec{a}|\approx0$, 
so that $\gamma \approx 1$. Condition (\ref{acceleration}) is immediately 
satisfied, the traveler feels a zero gravitational 
acceleration.

\bigskip\vskip1cm 
{\it (iii) Total time in a traversal} 
 
The expressions for the total times in a traversal of the 
wormhole, measured by the traveler and observers at rest at the 
station, are given by Equations 
(\ref{travelertime})-(\ref{observertime}). $l=-l_1$ and $l=l_2$ 
are the positions of the space stations. For low velocities, $v\ll 
c$, we have $\gamma \approx 1$, and with $\Phi=\Phi_0$, Equations 
(\ref{travelertime})-(\ref{observertime}) reduce to 
\begin{equation} 
\Delta \tau_{\rm traveler} \approx \frac{2l}{v}=e^{\Phi_0}\, 
\Delta t_{\rm space\,station}
\label{traversaltime1} 
\end{equation} 
Suppose that the space stations are placed in the neighborhood of 
the mouth, at $a$, in the exterior side. 
It is convenient to place the space stations at large 
enough radii, i.e., $a\gg r_{\rm o}$, so that the factor 
$1-b(r)/r\approx 1$. Thus, one also has that 
$e^{2\Phi(a)}=1-(r_{\rm o}/a)^{1/2}\approx 1$, so that 
$\Delta \tau_{\rm traveler} 
\approx \Delta t_{\rm space\,station}\approx 2a/v$. 
Assume that  the 
traversal time should be less than a year, i.e. $\Delta 
\tau_{\rm traveler} \leq 1\,{\rm year}$ 
($1$ year $=3.16\times 10^7 \,{\rm 
s}$). Therefore, from $2a/v \approx 1\,{\rm year}$, one extracts a 
value for $a$, namely $a\approx 4.74\times 10^{13}\,{\rm m}$.

\bigskip \bigskip
\noindent {\bf (2) The wormhole with shape function $b(r)=r_{\rm o}^2/r$} 
\medskip

Applying a similar analysis as before, Equation (\ref{tidal2}), 
with $v\ll c$ and $\gamma \approx 1$, and considering the form 
function given by $b(r)=r_{\rm o}^2/r$, we have 
\begin{equation} 
\left(\frac{v}{c}\right) \leq 
\left(\frac{r^2}{r_{\rm o}}\right)\frac{1}{10^8 {\rm m}} 
\end{equation} 
Imposing a traversal velocity 
$v=0.01\,c$
at the throat, $r=r_{\rm o}$, one finds $r_{\rm o} \approx 10^6 {\rm m}$. 
For $\Phi=\Phi_0$ the local 
gravitational accelerations at the space stations are zero. 
The traversal time, with $\gamma \approx 1$ is given 
by 
\begin{equation} 
\Delta \tau_{\rm traveler} \approx \frac{2l}{v}=e^{\Phi_0}\, 
\Delta t_{\rm space\,station} \,. 
\end{equation} 
By a similar analysis as above, we place the space 
stations at $a\gg r_{\rm o}$, implying that 
$e^{2\Phi(a)}=1-(r_{\rm o}/a)^2\approx 1$, so that $\Delta 
\tau_{\rm traveler}  
\approx \Delta t_{\rm space\,station}$. 
Considering that the traversal time is 
approximately one year, $\Delta \tau_{\rm traveler}  =
3.16\times 10^7 
{\rm s}$, and taking into account that the traversal velocity is 
$v=0.01\,c$, the junction surface is at $a=4.74\times 10^{13} 
{\rm m}$.

\end{document}